\documentclass[structabstract]{aa}  
\usepackage{graphicx}
\usepackage{txfonts}
\usepackage{natbib}
\bibpunct{(}{)}{;}{a}{}{,}
\usepackage{hyperref}
\defcitealias{r05}{Paper~I}
\usepackage{lscape}
\begin{document}
   \title{Quantifying the uncertainties of chemical evolution studies}

   \subtitle{II. Stellar yields}

   \author{D. Romano
          \inst{1, 2},
          A. I. Karakas
          \inst{3},
          M. Tosi
          \inst{2},
          \and
          F. Matteucci
          \inst{4, 5}
          }

   \institute{Dipartimento di Astronomia, Universit\`a di Bologna,
              Via Ranzani 1, I-40127 Bologna, Italy\\
               \email{donatella.romano@oabo.inaf.it}
         \and
             INAF, Osservatorio Astronomico di Bologna, 
             Via Ranzani 1, I-40127 Bologna, Italy\\
              \email{monica.tosi@oabo.inaf.it}
         \and
             Research School of Astronomy and Astrophysics, Mt. Stromlo 
             Observatory, Cotter Rd., Weston Creek, ACT 2611, Australia\\
              \email{akarakas@mso.anu.edu.au}
         \and
             Dipartimento di Fisica, Universit\`a di Trieste,
             Via Tiepolo 11, I-34143 Trieste, Italy\\
              \email{matteucc@oats.inaf.it}
         \and
             INAF, Osservatorio Astronomico di Trieste,
             Via Tiepolo 11, I-34143 Trieste, Italy\\
             }

   \date{Received 23 March 2010; accepted 28 June 2010}

 
  \abstract
   {Galactic chemical evolution models are useful tools to interpret the large 
     body of high-quality observational data on the chemical composition of 
     stars and gas in galaxies which have become available in recent years.}
   {This is the second paper of a series which aims at quantifying the 
     uncertainties in chemical evolution model predictions related to the 
     underlying model assumptions. Specifically, it deals with the 
     uncertainties due to the choice of the stellar yields.}
   {We adopt a widely used model for the chemical evolution of the Galaxy and 
     test the effects of changing the stellar nucleosynthesis prescriptions on 
     the predicted evolution of several chemical species. Up-to-date results 
     from stellar evolutionary models are taken into account carefully.}
   {We find that, except for a handful of elements whose nucleosynthesis in 
     stars is well understood by now, large uncertainties still affect the 
     model predictions. This is especially true for the majority of the 
     iron-peak elements, but also for much more abundant species such as carbon 
     and nitrogen. The main causes of the mismatch we find among the outputs of 
     different models assuming different stellar yields and among model 
     predictions and observations are: (i)~the adopted location of the mass cut 
     in models of type II supernova explosions; (ii)~the adopted strength and 
     extent of hot bottom burning in models of asymptotic giant branch stars; 
     (iii)~the neglection of the effects of rotation on the chemical 
     composition of the stellar surfaces; (iv)~the adopted rates of mass loss 
     and of (v)~nuclear reactions, and (vi)~the different treatments of 
     convection.}
   {Our results suggest that it is mandatory to include processes such as hot 
     bottom burning in intermediate-mass stars and rotation in stars of all 
     masses in accurate studies of stellar evolution and nucleosynthesis. In 
     spite of their importance, both these processes still have to be better 
     understood and characterized. As for massive stars, presupernova models 
     computed with mass loss and rotation are available in the literature, but 
     they still wait for a self-consistent coupling with the results of 
     explosive nucleosynthesis computations.}


   \titlerunning{Quantifying the uncertainties of chemical evolution studies. 
     II.}
   \authorrunning{D.~Romano et al.}

   \maketitle
%

\section{Introduction}
\label{sec:intro}

A large, everexpanding body of high-quality observational data is available 
nowadays for our own Galaxy, as well as for a few nearby galaxies, concerning 
the detailed chemical properties of their interstellar medium (ISM) and stellar 
populations. This wealth of data can be interpreted in the framework of 
galactic chemical evolution (GCE) models. Even though GCE is not yet a 
well-sound astrophysical theory, it is still a powerful instrument to 
understand when and how a given system formed and evolved. In fact, as 
different chemical elements are produced on different time scales in different 
astrophysical sites, the behaviour of key abundance ratios for a given galactic 
region allows us to trace the chronology of events in that region. Abundance 
ratios as a function of time or metallicity also provide important tests of 
possible nucleosynthesis scenarios \citep{t79}.

The main players in this game are the stellar yields, defined by \citet{t80} as 
the mass fractions of stars which are returned to the ISM in the form of newly 
produced elements during the entire stellar lifetimes. Following the pioneering 
works by \citet{it78} and \citet{rv81} for low- and intermediate-mass stars 
(LIMSs), and \citet{a78} and \citet{cc79} for massive stars, stellar models 
with an increasingly improved treatment of the relevant physical processes have 
appeared in the literature, thus providing different grids of yields. One of 
the most important differences among the various sets of yields is the presence 
of mass loss in massive star models. After \citeauthor{cc79}'s 
\citeyearpar{cc79} paper, the most important work on yields from massive stars 
to include mass loss was from \citet{m92}. \citet{m92} showed that mass loss 
driven by stellar winds mostly affects massive star models with solar or 
supersolar metallicities. In this case, stars lose large fractions of their 
newly produced helium and carbon. This has the effect of depressing the 
production of heavier elements, and in particular oxygen \citep[see 
also][]{d03,dt03}. This strong effect is particularly important in high 
metallicity environments such as the Galactic bulge \citep[see][]{mc08,c09}. In 
the following years the Geneva group has produced yields for massive stars with 
a lower mass loss rate than used previously by \citet{m92}, as well as the 
inclusion of stellar rotation \citep{mm02a,hmm05,h07}. The most significant 
result was the substantial quantity of primary N produced by including rotation 
in very-low metallicity massive star models \citep{h05,h07,c06}. Unfortunately, 
yields including mass loss through stellar winds are available only from a 
limited number of studies, whereas the majority of nucleosynthesis studies 
including a large number of chemical species do not include mass loss from 
massive stars. Among these we recall \citet{ww95,lc03,cl04}. The recent study 
by \citet{k06} includes mass loss but no rotation. Concerning LIMSs, 
nucleosynthesis yields including mass loss were from \citet{rv81}, 
\citet{vdhg}, \citet{m01}. More recently \citet{kl07} and \citet{k10} have 
provided metallicity dependent yields. Concerning type Ia supernovae (SNeIa, 
exploding white dwarfs in binary systems) the most commonly adopted yields are 
those from \citet[][their model W7]{n84}, subsequently revised by \citet{i99}. 
These yields are preferred over others because they can very well reproduce the 
observed abundance pattern in SNIa ejecta. Empirical yields for massive stars 
and SNIa were derived by \citet{f04}, who modified the existing yields in order 
to reproduce the solar abundances and the [X/Fe]\footnote{Throughout this 
paper, the standard notations [A/B]~$\equiv 
\log$~($N_{\mathrm{A}}$/$N_{\mathrm{B}}$)$-
\log$($N_{\mathrm{A}}$/$N_{\mathrm{B}}$)$_{\sun}$ and $\log 
\varepsilon$(A)~$\equiv \log$($N_{\mathrm{A}}$/$N_{\mathrm{H}}$)$+$12 are used, 
where $N_{\mathrm{A}}$, $N_{\mathrm{B}}$ and $N_{\mathrm{H}}$ are the number 
densities of elements A, B and hydrogen. Solar abundances are taken from 
\citet{gs98}, unless otherwise specified.} versus [Fe/H] patterns observed in 
the solar vicinity. Their yields mainly differ from the standard ones \citep[in 
this case those of][]{ww95} for Mg, Si, and the Fe-peak elements. However, it 
should be noted that no computed yields can reproduce the abundance patterns of 
the Fe-peak elements in the solar vicinity. 

The yields currently available for the most abundant elements result from 
updated stellar evolution and nucleosynthesis calculations, but differ quite 
significantly from each other. This implies that chemical evolution models 
assuming different yields can predict quite different element abundances even 
if they assume similar galactic parameters, such as the infall and star 
formation rate (SFR) or the initial mass function (IMF). It is therefore 
important to quantify what is the effect of different yields on the model 
results. To this purpose we compare stellar yields currently available in the 
literature and describe the corresponding differences in the outcome of a well 
tested chemical evolution model for our Galaxy. 

The layout of the paper is as follows. In Sect.~\ref{sec:yields} we concisely 
review the prescriptions about the physical ingredients of the stellar models 
adopted by different authors. This provides the different yield sets explored 
in this study. A comparison between the different yield sets is also presented.
In Sect.~\ref{sec:model} we introduce the adopted GCE model. Our results are 
presented and compared with the corresponding observational data in 
Sects.~\ref{sec:abrat} to \ref{sec:dydz}. In Sect.~\ref{sec:abrat} we deal with 
the evolution of elements from carbon to zinc in the solar neighbourhood. The 
Galactic abundance gradients are briefly discussed in Sect.~\ref{sec:grad}. The 
helium variation with metallicity is analysed in Sect.~\ref{sec:dydz}. 
Sect.~\ref{sec:conc} summarizes and discusses the results.


\section{Stellar yields}
\label{sec:yields}

In this section we provide a detailed comparison between the different yield 
sets adopted in this work. The reader is referred to the source papers for a 
more thorough discussion of the adopted input physics.

For LIMSs (see Sect.~\ref{sec:limsy}), we consider the yields by: (i) 
\citet{vdhg}, for two choices of the mass loss rate and minimum core mass for 
hot bottom burning; (ii) \citet{m01}, for two choices of the mixing-length 
parameter; (iii) \citet{kl07} standard set, taking the contribution from `extra 
pulses' into account, and (iv) \citet{k10}, who has recently recomputed 
\citeauthor{kl07}'s \citeyearpar{kl07} models with updated nuclear reaction 
rates and enlarged the grid of masses and metallicities.

For massive stars (see Sect.~\ref{sec:msy}), we consider the yields by: (i) 
\citet{ww95}, computed without mass loss and without rotation; (ii) 
\citet{k06}, including metallicity-dependent mass loss, and (iii) the Geneva 
group \citep{mm02a,hmm05,h07,e08}, limited to the presupernova stage, but 
computed with both mass loss and rotation. Older yields by \citet{m92} for 
solar-metallicity stars -- with higher mass loss rates but without rotation -- 
are considered as well. All of the adopted yield sets are metallicity dependent 
and are provided for various metallicities.

\subsection{Low- and intermediate-mass stars}
\label{sec:limsy}

Stars with initial masses between $\sim$1 and $\sim$5--8 M$_{\sun}$ (depending 
on the overshooting adopted in the stellar evolutionary models) pass through a 
phase of double-shell burning, referred to as thermally-pulsing asymptotic 
giant branch (TP-AGB) phase, before ending their lives as white dwarfs. During 
the TP-AGB phase, they experience a number of thermal pulses followed by mixing 
episodes, the third dredge-up (TDU). In the most massive models, hot bottom 
burning (HBB) can occur if the base of the convective envelope is exposed to 
temperatures hot enough to trigger proton-capture nucleosynthesis. During the 
TP-AGB evolution, the stars experience a very rich nucleosynthesis whose 
products (most notably He, C, N, Na, Mg and \emph{s-}process elements) are 
convected to the outermost layers and eventually injected into the ISM by 
stellar winds. These stars are hence important contributors to the chemical 
enrichment of galaxies.

\citet{vdhg} published yields of \element[][4]{He}, \element[][12]{C}, 
\element[][13]{C}, \element[][14]{N} and \element[][16]{O} for stars with 
initial masses in the range 0.8--8 M$_{\sun}$ and initial compositions ($Y$, 
$Z$)~= (0.243, 0.001), (0.252, 0.004), (0.264, 0.008), (0.30, 0.02) and (0.34, 
0.04). They used the metallicity-dependent tracks provided by the Geneva group 
\citep[e.g.][]{s92} to follow the stellar evolution prior to the AGB phase, 
with the recipes outlined in \citet{gj93} to include the effects of the first 
and second dredge-ups on the yields. Then, they followed the chemical evolution 
and mass loss up to the planetary nebula (PN) ejection with a synthetic TP-AGB 
evolution model. The effect of HBB was included in an approximate way and a 
\citet{r75} law was assumed for the mass loss of stars on the AGB. These models 
do not consider any extra mixing. In this work we explore the \citet{vdhg} 
yield sets with mass loss scaling parameter along the AGB, $\eta_\mathrm{AGB}$, 
either varying with metallicity or constant. When the set with a constant 
$\eta_\mathrm{AGB}$ is assumed, we also investigate the effect of reducing the 
mass range in which HBB is allowed to occur.

\citet{m01} followed the evolution of LIMSs with initial mass from $\sim$0.8 to 
5~M$_{\sun}$ for three choices of the original composition, namely ($Y$, $Z$)~= 
(0.240, 0.004), (0.250, 0.008) and (0.273, 0.019). From the zero-age main 
sequence up to the onset of the TP-AGB phase, she has used complete stellar 
models by \citet{g00}, including moderate core overshooting and external 
convection, but no extra mixing. Later phases have been treated with the help 
of synthetic TP-AGB models \citep{mgb99}. The classical \citet{r75} formula has 
been used to include mass loss along the red giant branch (RGB), while mass 
loss during the TP-AGB phase has been included by means of the semi-empirical 
formalism developed by \citet{vw93}. Yields of \element[][3]{He}, 
\element[][4]{He}, \element[][12]{C}, \element[][13]{C}, \element[][14]{N}, 
\element[][15]{N}, \element[][16]{O}, \element[][17]{O} and \element[][18]{O} 
are provided for three choices of the mixing-length parameter, namely 
$\alpha$~= 1.68, 2.00 and 2.50. We investigate the effect of choosing the 
yields corresponding to either the lowest or the highest $\alpha$ value in 
Sect.~\ref{sec:cno}.

\begin{table*}
\caption{Yields for massive stars from presupernova Geneva models including 
mass loss and rotation. Listed are: the initial metallicity of the stars, 
$Z_{\mathrm{ini}}$, the initial stellar masses (in M$_{\sun}$) and rotational 
velocities (in km s$^{-1}$), $m_{\mathrm{ini}}$($\upsilon_{\mathrm{ini}}$), and 
the reference papers from which the yields have been picked up. See the text 
for details.}
\label{tab:rotyld}
\centering
\renewcommand{\footnoterule}{}  
\begin{tabular}{c l c}
\hline \hline
$Z_{\mathrm{ini}}$ & $m_{\mathrm{ini}}$($\upsilon_{\mathrm{ini}}$) & Ref.\\
\hline
        0 & 9(500), 15(800), 25(800), 40(800), 60(800), 85(800) & 1\\
10$^{-8}$ & 9(500), 20(600), 40(700), 60(800), 85(800) & 2\\
10$^{-5}$ & 9(300), 15(300), 20(300), 40(300), 60(300) & 3\\
    0.004 & 9(300), 12(300), 15(300), 20(300), 25(300), 40(300), 60(0) & 3\\
     0.02 & 12(300), 15(300), 20(300), 25(300), 40(300), 60(300) & 4\\
\hline
\end{tabular}

\medskip
\begin{minipage}[]{10cm}
References. (1) \citet{e08}; (2) \citet{h07}; (3) \citet{mm02a}; (4) 
\citet{hmm05}.\\
\end{minipage}
\end{table*}

The AGB yields of \citet{kl07} and \citet{k10} were calculated from detailed 
stellar models, where the structure was computed first and the nucleosynthesis 
calculated afterward using a post-processing algorithm. Yields are included for 
77 nuclei including all stable isotopes from H to $^{34}$S, and for a small 
group of Fe-peak nuclei. The details of this procedure and the codes used to 
compute the models have been previously described in some detail, see for 
example \citet{k09} and references therein. The models cover a range in mass 
from 1.0~M$_{\sun}$ to 6~M$_{\sun}$ and compositions $Z$~= 0.02, 0.008, 0.004, 
and $Z$~= 0.0001. All models were evolved from the zero-age main sequence to 
near the tip of the TP-AGB. The TDU efficiency governs the nucleosynthesis in 
the lower mass models; this was found to vary as a function of the H-exhausted 
core mass, metallicity, and envelope mass \citep[see][for details]{k02}. For 
example, in the $Z$~= 0.02 models, no TDU was found for $m \le$~2 M$_{\sun}$. 
For the intermediate-mass models, the TDU was found to be efficient and the o
ccurrence of HBB also played a strong role in determining the final yields. The 
occurrence of HBB also depends on the initial mass and metallicity, with HBB 
occurring in lower mass stars with a decrease in metallicity (at 3~M$_{\sun}$ 
at $Z$~= 10$^{-4}$, whereas it only starts at 5~M$_{\sun}$ at $Z$~= 0.02). 
Furthermore, HBB is eventually shut down by the action of mass loss. For 
further details we refer to \citet{kl07} and \citet{k10}. These models employ 
the mixing-length theory of convection with $\alpha$~= 1.75. On the RGB 
Reimer's mass loss is adopted with $\eta_{\mathrm{R}}$~= 0.4. On the AGB, 
\citet{vw93} mass loss is used for all stellar models in \citet{kl07} and most 
models in \citet{k10}; in \citet{k10} Reimer's was used on the AGB for models 
between 3 to 6~M$_{\sun}$ with $Z$~= 10$^{-4}$ (see reference for details). The 
main difference between the \citet{kl07} models and those presented in 
\citet{k10} is the choice of reaction rates used in the post-processing 
algorithm. \citet{k10} used an updated set of proton and $\alpha$-capture rates 
that include some of the latest experimental results for important reactions 
involved in the CNO cycle, and the NeNa and MgAl chains. Furthermore, 
\citet{k10} assumed scaled-solar initial abundances for the $Z$~= 0.008 and 
$Z$~= 0.004 models: \citet{kl07} adopted initial abundances for the Large and 
Small Magellanic Clouds from \citet{rd92} which are sub-solar for C, N, and O. 
The main difference between the yields of \citet{kl07} and \citet{k10} is that 
$\sim$6--30 times less Na is produced by intermediate-mass models with HBB. The 
yields of \citet{kl07} also included the contribution from `extra pulses', that 
is thermal pulses that were not computed using the detailed stellar evolution 
code but with the help of a synthetic TP-AGB model. This was done because it 
was not always possible to evolve the models to the tip of the AGB owing to 
convergence difficulties. If there was sufficient envelope mass left, further 
thermal pulses and TDU could potentially occur. In \citet{k10}, the 
contribution from these extra pulses was \emph{not} included. This was because 
the synthetic AGB models required an assumption about the efficiency of the TDU 
for these last pulses. Full calculations have shown that the efficiency 
approaches zero as the envelope mass is decreased \citep{vw93,str97,k02}. The 
new intermediate-mass $Z$~= 0.0001 AGB models presented in \citet{k10} were 
evolved to small envelope masses ($\sim$0.1~M$_{\sun}$ in some cases). These 
models did not show any decrease in TDU efficiency with decreasing envelope 
mass, but in general low-metallicity models tend to experience more efficient 
mixing \citep[e.g.][]{bs88,k02}.

\subsection{Massive stars}
\label{sec:msy}

Only a small fraction of the stellar mass, from 7 to 20 per cent, depending on 
the choice of the IMF \citep[see][hereinafter Paper~I]{r05}, is in stars with 
masses above 8~M$_{\sun}$. Nevertheless, these stars are major galactic 
polluters. They produce almost the whole body of the $\alpha$ and $r$-process 
elements and, if rotating fast, large amounts of primary \element[][13]C, 
\element[][14]N, \element[][22]Ne and $s$-process elements at low metallicities 
\citep[see][for a review on rotating high-mass stars]{m09}. Massive stars are 
also thought to be the main factories of copper in galaxies \citep{rm07}.

\citet{ww95} provided the nucleosynthetic yields of isotopes lighter than $A$~= 
66 (zinc) for stars of various metallicities ($Z/Z_{\sun}$~= 0, 0.0001, 0.01, 
0.1 and 1) and masses in the range 11--40 M$_{\sun}$. In their models, each 
star is exploded using a piston to give a final kinetic energy of the ejecta at 
infinity of about 10$^{51}$ ergs. The explosive modifications to the 
presupernova nucleosynthesis are computed by taking the effects of neutrino 
irradiation into account. Mass loss is not included in the presupernova models. 
Following the onset of a successful shock, stars more massive than 
30~M$_{\sun}$ may experience considerable reimplosion of heavy elements. The 
amount of matter which falls back onto the collapsed remnant depends mainly on 
the initial energy of the explosion. For stars with $m \ge$~30 M$_{\sun}$, 
yields are given for different choices of the explosion energy \citep[models 
labelled A, B, C in][]{ww95}. We adopt both case A and case B, the latter 
corresponding to a slightly larger final kinetic energy of the ejecta at 
infinity (typically 1.9~$\times$~10$^{51}$ ergs rather than 
1.2~$\times$~10$^{51}$ ergs). We systematically halve their Fe yields. This 
leads to a better fit to most of the observed abundance ratios (see 
Sect.~\ref{sec:abrat}). A factor of 2 of reduction is well within the 
uncertainties in modelling the explosion. Furthermore, the reduced Fe yields 
are consistent with observations of several type II supernovae 
\citep[SNeII;][]{t95}.

It is now acknowledged that there exist two types of core-collapse SNe. One is 
normal SNeII (including Ib and Ic), with explosion energies of the order of 
10$^{51}$ ergs. The other is hypernovae (HNe), with more than 10 times larger 
explosion energies \citep{i98}. \citet{k06} have calculated the yields for both 
SNeII and HNe as functions of the initial masses (from 13 to 40 M$_{\sun}$) and 
metallicities ($Z$~= 0, 0.001, 0.004, 0.02) of the progenitor stars. Their 
calculations start on the main sequence and proceed through core collapse, with 
the inclusion of metallicity-dependent mass loss \citep{dj88,k89}. The 
progenitor model is exploded when a central density of 3~$\times$~10$^{10}$ g 
cm$^{-3}$ is reached. For SNeII, the final yields are obtained by setting the 
mass cut in such a way that 0.07~M$_{\sun}$ of Fe are ejected. For HNe, the 
final yields are obtained by adjusting the free parameters involved in the 
`mixing and fallback' mechanism \citep{un02} to give [O/Fe]~$\simeq$~0.5 in the 
ejecta, as observed in extremely metal-poor stars. In order to include HN 
nucleosynthesis in GCE models, one should introduce a free parameter describing 
the fraction of HNe. In the present work, we analyze the two extreme cases 
where either none ($\varepsilon_{\mathrm{HN}}$~= 0) or all 
($\varepsilon_{\mathrm{HN}}$~= 1) stars with $m \ge$~20 M$_{\sun}$ explode as 
HNe.

The inclusion of rotation in stellar evolutionary models considerably alters 
the outputs of traditional nucleosynthesis. Yields of \element[][4]{He}, 
\element[][12]{C}, \element[][14]{N} and \element[][16]{O} are available for 
high-mass rotating stars, for a fine grid of initial masses and metallicities 
\citep[][see Table~\ref{tab:rotyld}]{mm02a,hmm05,h07,e08}. Yields of 
\element[][13]{C}, \element[][17]{O} and \element[][18]{O} are provided only 
for a subset of models. The models include diffusion by shears, meridional 
circulation and mass loss and are computed up to the end of the C-burning 
phase. The absence of a proper, self-consistent computation of the 
nucleosynthesis occurring at the SN stage is the major shortcoming of such 
models. However, the presupernova yields of \element[][4]{He} and CNO nuclei 
should be almost unaffected by subsequent explosive nucleosynthesis 
\citep[e.g.][]{ww95} and can thus be safely used in GCE models. The grid of 
yields for high-mass rotators that we use in this study has been obtained by 
patching together results presented in different papers by the same group (see 
Table~\ref{tab:rotyld}). We also test the models computed by \citet{mm02a}, 
\citet{hmm05} and \citet{e08} for non-rotating high-mass stars (see 
Sect.~\ref{sec:cno}).

\subsection{Comparison among different yield sets}
\label{sec:yldcomp}

\begin{figure*}
  \centering
  \includegraphics[width=\textwidth]{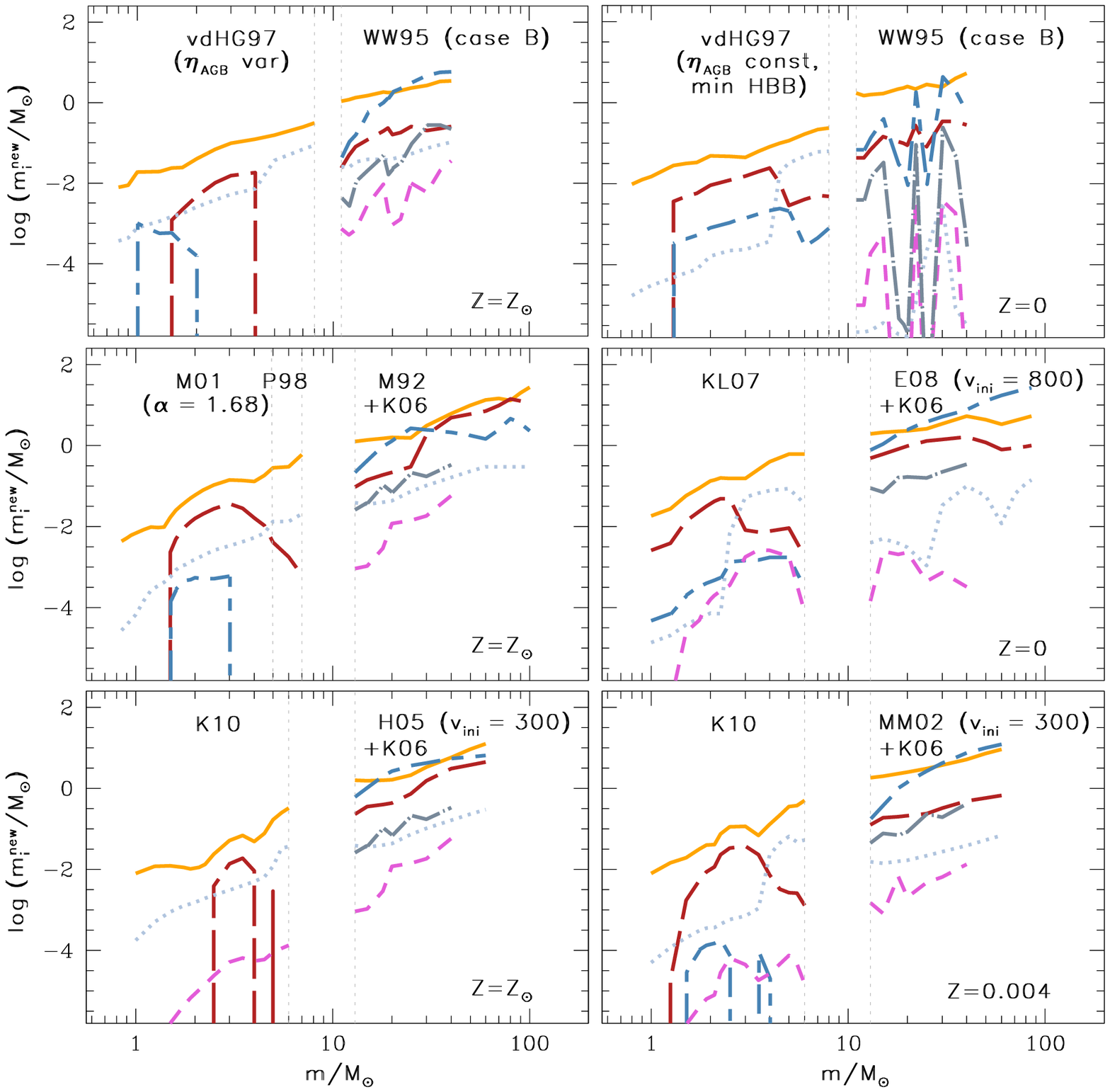}
  \caption{ Ejected masses of newly produced elements as a function of the 
    stellar initial mass, from different authors (vdHG97: \citealt{vdhg}; WW95: 
    \citealt{ww95}; M01: \citealt{m01}; P98: \citealt{p98}; M92: \citealt{m92}; 
    K06: \citealt{k06}; KL07: \citealt{kl07}; E08: \citealt{e08}; K10: 
    \citealt{k10}; H05: \citealt{hmm05}; MM02: \citealt{mm02a}) and for 
    different initial metallicities. Solid (orange) lines: He; long-dashed 
    (red) lines: C; dotted (light blue) lines: N; short-dashed-long-dashed 
    (blue) lines: O; short-dashed (magenta) lines: Na; dot-dashed (gray) lines: 
    Mg. See the electronic edition of the journal for a colour version of this 
    figure.}
  \label{fig:yields}
\end{figure*}
%

\begin{figure*}
  \centering
  \includegraphics[width=\textwidth]{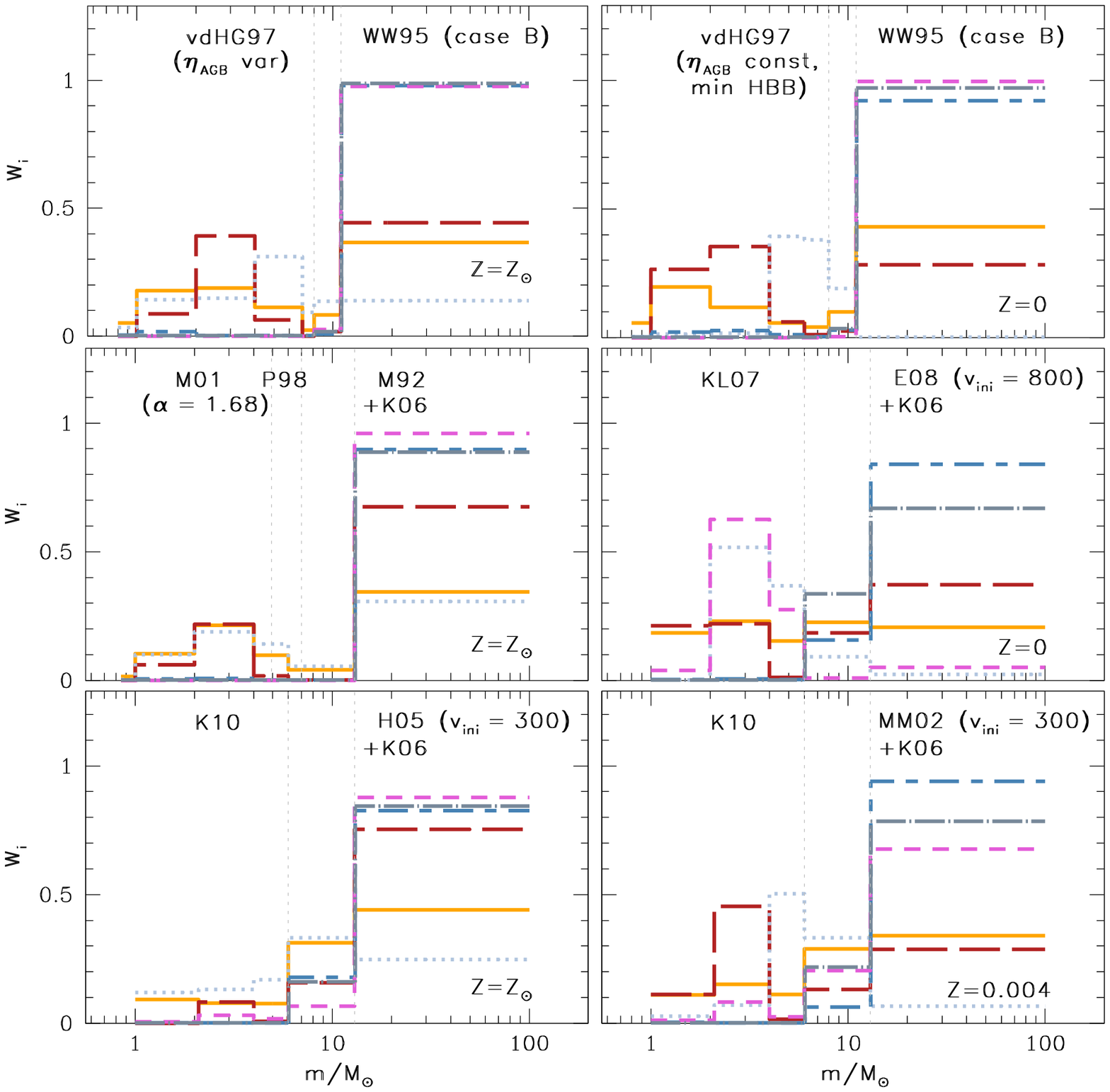}
  \caption{ IMF weighted yields in specific mass ranges. Line types and colours 
    are the same as in Fig.~\ref{fig:yields}. See the electronic edition of the 
    journal for a colour version of this figure.}
  \label{fig:imfw}
\end{figure*}
%


In the previous sections, we have considered a number of stellar evolution and 
nucleosynthesis studies. We have reviewed the basic input physics used in the 
stellar structure computations and schematically discussed the resulting yield 
sets. Other works on stellar evolution and nucleosynthesis exist in the 
literature besides the ones considered here, providing different grids of 
yields. However, these grids are computed either for mass/metallicity ranges 
smaller than the ones used here, or without the inclusion of some important 
physical processes affecting the final yields. Others are not much different 
from those considered here. The latter, for instance, is the case of the yields 
presented by \citet{iz04}, which are very similar to the ones by \citet{kl07} 
(see figures from 2 to 7 of \citealt{kl07}) and are thus not included in the 
present study. The Geneva group provides yields for LIMSs to be used to 
complete their grid of high-mass models. However, the computation of the LIMS 
models is stopped at the beginning of the TP-AGB phase. When dealing with 
LIMSs, the inclusion of the final nuclear burning phase is an essential step 
towards computing trustworthy yields of the CNO elements. Therefore, we prefer 
not to include the Geneva yields for LIMSs in our GCE models until they do 
cover the TP-AGB phase too. As a last example, \citet{vd08a,vd08b} have 
recently provided new grids of yields for intermediate-mass stars for many 
chemical species. With the main goal of testing the self-enrichment scenario by 
massive AGB stars for Galactic globular clusters \citep[see][and references 
therein]{vd08a}, their prescriptions cover a relatively narrow range of initial 
stellar masses (3--6.5~M$_{\sun}$) and chemical compositions ($Z$~= 0.001 and 
0.004). As such, they are not suitable for use in chemical evolution models of 
the Milky Way and other systems where stars of all masses and metallicity have 
contributed to the enrichment.

The results from stellar nucleosynthesis are steadily improving with time, but 
we are still missing a complete and homogeneous set of stellar yields -- the 
`ideal grid' of yields computed with the same input physics for various initial 
masses and metallicities of the stars, spanning the whole range from 0.8 to 
100~M$_{\sun}$ in initial stellar mass and from zero to (at least) twice solar 
in initial metallicity. Such an optimal yield set should also take into account 
all the relevant physical processes occurring in the stellar interiors and 
affecting the chemical composition of the ejected layers during the whole 
stellar lifetime \citep[see][for a recent appraisal of the problem]{t07}.

The potential risks of patching together different yield sets are highlighted 
in Figs.~\ref{fig:yields} and \ref{fig:imfw} \citep[see also][]{t07}. The 
left-hand panels of Fig.~\ref{fig:yields} show the solar yields of \citet{vdhg} 
for LIMSs combined with those of \citet{ww95} for massive stars (upper panel), 
the solar yields of \citet{m01} for LIMSs combined with those of \citet{m92} 
for He, C, N, O and \citet{k06} for all other elements from massive stars 
(middle panel) and the solar yields of \citet{k10} for LIMSs combined with 
those of \citet{hmm05} for He, C, N, O and \citet{k06} for all other elements 
from massive stars (lower panel). The \citet{m01} yield set is completed with 
the homogeneous yields by \citet{p98} for stars of initial mass $m$~= 
6~M$_{\sun}$ and $m$~= 7~M$_{\sun}$. Other combinations of yields for different 
metallicities\footnote{Since neither \citet{vdhg} nor \citet{kl07} provide 
yields for zero-metallicity stars, at $Z$~= 0 we adopt the yields corresponding 
to the lowest metallicities considered by those authors, namely $Z$~= 0.001 for 
\citet{vdhg} and $Z$~= 0.0001 for \citet{kl07}.} are shown in the right-hand 
panels of Fig.~\ref{fig:yields}.

Two problems are immediately apparent. First, a gap is present between the LIMS 
and high-mass star domains, from 8 to 11~M$_{\sun}$ or from 6 to 13~M$_{\sun}$, 
depending on the choice of the yield sets. This gap forces us to interpolate 
the yields over a non-negligible mass interval. This mass interval is the 
domain of the so-called super-AGB stars, that have not reliable nucleosynthetic 
output \citep[][and references therein]{s07,pdv08}. Moreover, as shown below, 
once weighted with the IMF this mass range is one of the major contributors to 
the Galactic enrichment. Second, several nucleosynthesis computations do not go 
beyond 40~M$_{\sun}$, which requires an uncertain extrapolation to higher 
masses. While, thanks to the negative slope of the IMF, this may be of little 
concern in modelling the recent chemical evolution of the Milky Way, it becomes 
extremely relevant for the very early epochs, when only the most massive stars 
were polluting the ISM.

To evaluate the contribution to the Galaxy enrichment of a given mass range, 
one must take the IMF into account. An efficient way to do that is to compute 
the IMF weighted yields. Let us first define for each element the integrated 
yield as:
\begin{equation}
P_i \equiv \int^{m_u}_{m_l} m_i^{\mathrm{new}}(m) \phi(m) dm,
\end{equation}
\label{eq:intyld}
where $m_l$~= 0.1~M$_{\sun}$ and $m_u$~= 100~M$_{\sun}$ are, respectively, the 
lower and upper mass limits to the mass range of all stars formed in the 
examined region; $m_i^{\mathrm{new}}(m)$ is the yield (in solar masses) and 
$\phi(m)$ is the IMF. The IMF weighted yield is:
\begin{equation}
W_i \equiv \frac{\int^{m_2}_{m_1} m_i^{\mathrm{new}}(m) \phi(m) dm}{P_i},
\end{equation}
\label{eq:wyld}
where $m_1$ and $m_2$ are the lower and upper mass limits of the examined range.

\begin{table*}
\caption{Nucleosynthesis prescriptions.}             
\label{tab:nucp}      
\centering
\renewcommand{\footnoterule}{}  
\begin{tabular}{l c c c}
\hline \hline
Model & \multicolumn{2}{c}{Adopted stellar yields} & Comments \\
 & LIMSs & Massive stars & \\
\hline
 1 & vdHG97, $\eta_\mathrm{AGB}$ var & WW95, case B & Reference model\\
 2 & vdHG97, $\eta_\mathrm{AGB}$ var & WW95, case A & Mass cut changed\\
 3 & vdHG97, $\eta_\mathrm{AGB}$ var & WW95, case B + M92 pre-SN yields & Winds from $Z = Z_{\sun}$ massive stars included\\
 4 & vdHG97, $\eta_\mathrm{AGB}$ var & K06, $\varepsilon_{\mathrm{HN}}$~= 0 & SNII yields changed\\
 5 & vdHG97, $\eta_\mathrm{AGB}$ var & K06, $\varepsilon_{\mathrm{HN}}$~= 1 & HN nucleosynthesis included\\
 6 & vdHG97, $\eta_\mathrm{AGB}$ var & K06, $\varepsilon_{\mathrm{HN}}$~= 1 + Geneva pre-SN yields, $\upsilon_{\mathrm{ini}} \neq$ 0 & Stellar rotation included\\
 7 & vdHG97, $\eta_\mathrm{AGB}$ var & K06, $\varepsilon_{\mathrm{HN}}$~= 1 + Geneva pre-SN yields, $\upsilon_{\mathrm{ini}}$~= 0 & \\
 8 & vdHG97, $\eta_\mathrm{AGB}$ const & K06, $\varepsilon_{\mathrm{HN}}$~= 1 + Geneva pre-SN yields, $\upsilon_{\mathrm{ini}} \neq$ 0 & Mass loss along the AGB changed\\
 9 & vdHG97, minimum HBB & K06, $\varepsilon_{\mathrm{HN}}$~= 1 + Geneva pre-SN yields, $\upsilon_{\mathrm{ini}} \neq$ 0 & HBB extent reduced\\
 10 & M01, $\alpha$~= 1.68 & K06, $\varepsilon_{\mathrm{HN}}$~= 1 + Geneva pre-SN yields, $\upsilon_{\mathrm{ini}} \neq$ 0 & LIMS yields changed\\
 11 & M01, $\alpha$~= 2.50 & K06, $\varepsilon_{\mathrm{HN}}$~= 1 + Geneva pre-SN yields, $\upsilon_{\mathrm{ini}} \neq$ 0 & HBB strength increased\\
 12 & M01, $\alpha$~= 1.68 & K06, $\varepsilon_{\mathrm{HN}}$~= 1 + [Geneva, $\upsilon_{\mathrm{ini}} \neq$ 0 + M92] pre-SN yields & \\
 13 & KL07, with extra pulses & K06, $\varepsilon_{\mathrm{HN}}$~= 1 + Geneva pre-SN yields, $\upsilon_{\mathrm{ini}} \neq$ 0 & AGB yields from detailed stellar models\\
 14 & KL07, with extra pulses & K06, $\varepsilon_{\mathrm{HN}}$~= 1 + [Geneva, $\upsilon_{\mathrm{ini}} \neq$ 0 + M92] pre-SN yields & \\
 15 & K10, without extra pulses & K06, $\varepsilon_{\mathrm{HN}}$~= 1 + Geneva pre-SN yields, $\upsilon_{\mathrm{ini}} \neq$ 0 & Up-to-date nuclear reaction rates for LIMSs\\
\hline
\end{tabular}
%

\medskip
Notes. vdHG97: \citet{vdhg}; WW95: \citet{ww95}; M92: \citet{m92}; K06: 
\citet{k06}; M01: \citet{m01}; KL07: \citet{kl07}; K10: \citet{k10}. The models 
adopting the yields by \citet{m01} for LIMSs use self-consistent 
nucleosynthesis prescriptions from \citet{p98} for stars of initial mass $m$~= 
6~M$_{\sun}$ and $m$~= 7~M$_{\sun}$.\\
\end{table*}

In Fig.~\ref{fig:imfw}, the yields of Fig.~\ref{fig:yields} are weighted with 
the IMF of \citet{k93}. In computing the weighted yields, a linear 
interpolation is performed over the 8--11~M$_{\sun}$ (6--13~M$_{\sun}$, 
depending on the stellar models) mass range. When needed, the yields for 
massive stars are extrapolated to 100~M$_{\sun}$ by keeping constant the last 
computed masses ejected as newly produced elements. Very often, a bump is seen 
in the production of specific elements from stars in the 8--11 
(6--13)~M$_{\sun}$ interpolated mass range. In practice, more than 30\% of all 
He and N newly produced by a stellar generation can come from stars in the 
6--13~M$_{\sun}$ interval if the \citet{k10}+Geneva+\citet{k06} combination of 
yields is adopted (Fig.~\ref{fig:imfw}, bottom panels). This enhanced 
contribution can be completely spurious.

The only way to overcome these problems is getting homogeneous and complete 
sets of yields.

\section{The chemical evolution model}
\label{sec:model}

To compare the contribution to the Galaxy enrichment of the yields described in 
the previous sections, we adopt the \emph{two-infall model} case B for the 
chemical evolution of the Galaxy. A thorough discussion of the adopted 
formalism and basic equations can be found elsewhere \citep{c97,c01}. Here we 
briefly recall the overall evolutionary scenario.

The inner halo and thick disc of the Milky Way are assumed to form on a 
relatively short time-scale (about 1~Gyr) out of a first infall episode, 
whereas the thin disc forms inside-out \citep{mf89} on longer time-scales 
\citep[7~Gyr in the solar vicinity and more than a Hubble time at the outermost 
radii; see also][]{r00} during a second independent episode of extragalactic 
gas infall. The Galactic disc is approximated by several independent rings, 
2~kpc wide. Radial flows are not considered here. Galactic outflows seem 
unlikely owing to the deep Galactic potential well, and are thus not included 
in the model. The adopted SFR is proportional to both the total mass and the 
gas surface densities. The efficiency of conversion of gas into stars is higher 
during the halo/thick-disc phase than during the thin-disc phase. Furthermore, 
it drops to zero every time the gas density drops below a critical density 
threshold. The instantaneous recycling approximation is relaxed, i.e. the 
stellar lifetimes are taken into account in details. As for the stellar IMF, 
the \citet{k93} IMF is assumed \citepalias[see][and further discussion in 
Sect.~\ref{sec:dydz}]{r05} in the 0.1--100~M$_{\sun}$ mass range. The rate of 
SNIa explosions is calculated as in \citet{mg86}. SNeIa explode in close binary 
systems when a CO white dwarf has reached a critical mass limit because of 
accretion of hydrogen-rich matter from a main sequence or red giant companion.

The yields for SNeIa are taken from \citet{i99}, and we use their model~W7. For 
single stars, we adopt different sets of stellar yields (see 
Table~\ref{tab:nucp} and discussion in Sect.~\ref{sec:yields}). Overall, we 
compute 15 models, which differ only in the adopted nucleosynthesis 
prescriptions. Another couple of models, Models~1S and 10S, have been computed 
with different assumptions on the IMF, to let the reader appreciate the IMF 
effect on the results. They are the same as Models~1 and 10, respectively, but 
assume a \citeauthor{s86}'s \citeyearpar{s86} IMF rather than a 
\citeauthor{k93}'s \citeyearpar{k93} IMF. They are discussed at length in 
Sect.~\ref{sec:dydz} and are not listed in Table~\ref{tab:nucp}. A linear 
interpolation of the yields is performed as a function of both mass and 
metallicity. Since none of the massive star studies described in the previous 
sections includes prescriptions for a 100-M$_{\sun}$ star (see 
Sect.~\ref{sec:msy}), we must extrapolate the yields available for the most 
massive stellar models. We choose to keep the masses ejected in form of newly 
produced elements fixed up to 100~M$_{\sun}$.

In the following sections, we present the model predictions and compare them to 
one another and with the relevant data.


\section{Results on the abundance ratios}
\label{sec:abrat}

In this section we discuss, element by element, the behaviour of several 
abundance ratios as a function of metallicity in the solar neighbourhood, [X/M] 
versus [M/H]. We show either iron, oxygen or magnesium as the metallicity 
indicator.

\subsection{CNO nuclei}
\label{sec:cno}

\paragraph{\bf Carbon} 
Carbon is known to form through the 3-$\alpha$ process in stars \citep[][and 
references therein]{wal97}. However, it is still debated whether it comes 
mainly from LIMSs or massive stars \citep[see e.g.][for different points of 
view]{h00,crm03,ake04,car05,c09}.

\begin{figure*}
  \centering
  \includegraphics[width=\textwidth]{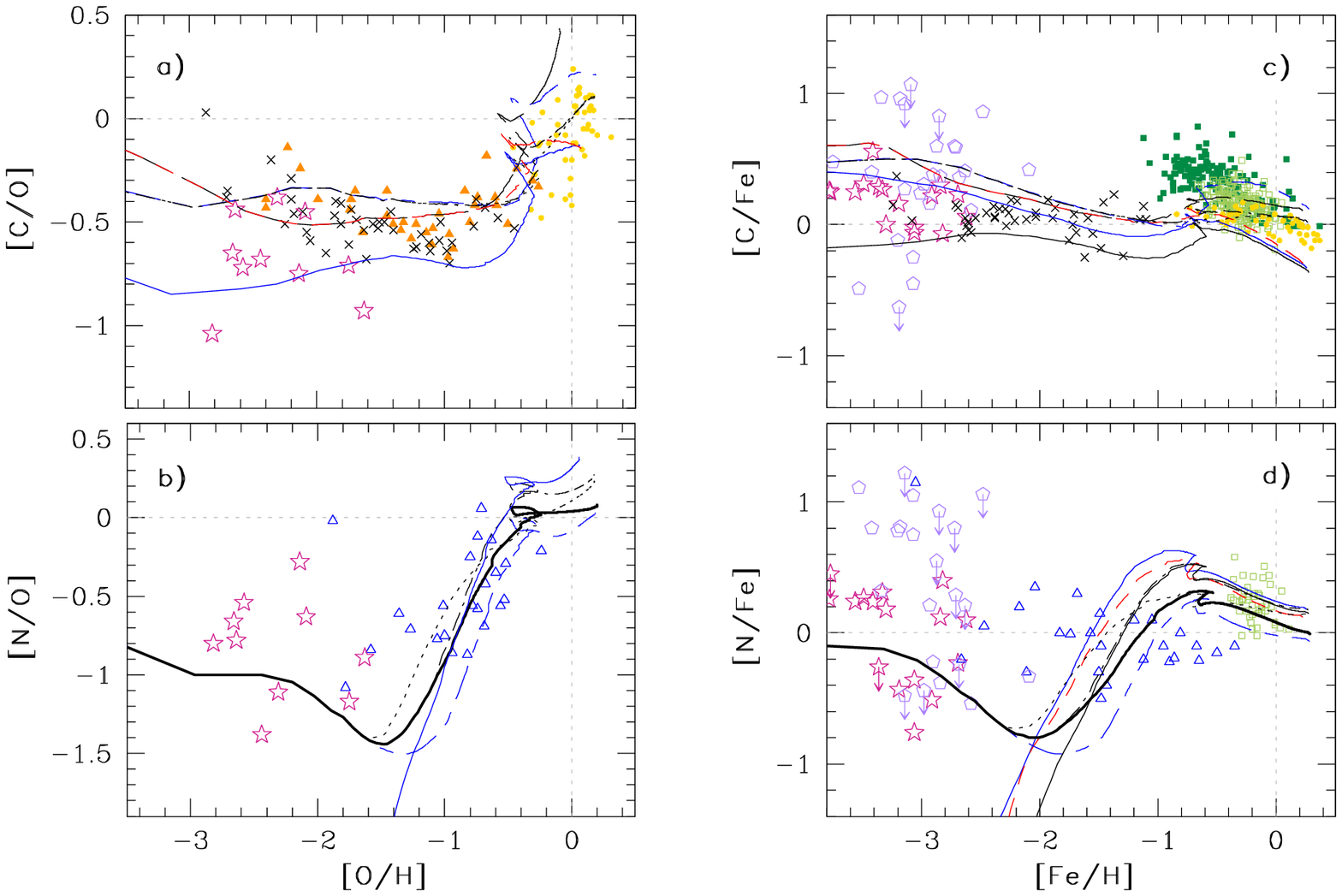}
  \caption{Left panels: Carbon-to-oxygen (upper panel) and nitrogen-to-oxygen 
    (lower panel) ratios as functions of [O/H] in the solar neighbourhood. 
    Right panels: [C/Fe] (upper panel) and [N/Fe] (lower panel) as functions of 
    [Fe/H] in the solar neighbourhood. Data are from \citet[][empty 
    stars]{s05}, \citet[][empty pentagons]{l08}, \citet[][crosses]{f09}, 
    \citet[][filled tringles]{ake04}, \citet[][open triangles]{i04}, 
    \citet[][filled squares]{rlap06} \citet[][open squares]{r03} and 
    \citet[][filled circles]{bf06}. Model predictions are shown for Models~1 
    [short-dashed (red) lines], 3 [long-dashed (black) lines], 4 [solid (blue) 
    lines], 5 [solid (black) lines], 6 [dot-dashed (black) lines], 9 [dotted 
    (black) lines], 10 [short-dashed-long-dashed (blue) lines] and 15 (thick 
    solid lines). See the electronic edition of the journal for a colour 
    version of this figure.}
  \label{fig:cn}
\end{figure*}

In Fig.~\ref{fig:cn}, we show the behaviour of [C/O] as a function of [O/H] 
(panel a) and [C/Fe] as a function of [Fe/H] (panel c) as traced by 
high-resolution observations of solar neighbourhood stars (symbols). The data 
are compared to the predictions of selected models (lines).

In order to derive the stellar carbon abundances, spectroscopists rely on 
high-excitation permitted \ion{C}{I} features and on the forbidden [\ion{C}{I}] 
line at 872.7 nm (free from non-LTE effects), as well as on various molecular 
species, notably CH, C$_2$ and CO. For solar-type stars, carbon abundances 
based on the forbidden [\ion{C}{I}] line at 872.7 nm are highly accurate, 
especially if the blending weak feature affecting the line is taken into 
account \citep{bf06}. According to \citet{bf06}, who analyzed the forbidden 
[\ion{C}{I}] line in a sample of 51 nearby F and G dwarf stars, [C/Fe] versus 
[Fe/H] is flat for subsolar metallicities down to [Fe/H]~= $-$1.0, with no 
distinctive trends for thin- and thick-disc stars. At higher metallicities, 
from [Fe/H]~$\approx$ 0 and up to [Fe/H]~$\approx$ +0.4, thin-disc members 
(which extend to higher metallicities than thick-disc members) show a shallow 
decline in [C/Fe]. The [C/O] versus [O/H] trends instead are well separated 
between the two discs, thanks to differences in the oxygen abundances. On the 
other hand, \citet{rlap06}, using permitted \ion{C}{I} lines, find that the 
abundance ratios [C/Fe] for the thick disc are on average larger than the 
[C/Fe] for the thin disc in the metallicity range $-$1.0~$<$ [Fe/H]~$<$ $-$0.4. 
Furthermore, [C/Fe] is found to increase with decreasing [Fe/H] at sub-solar 
metallicities. A trend of increasing [C/Fe] with decreasing [Fe/H] results also 
from the metal-poor halo stars analyzed by \citet{ake04}. In the [C/O] versus 
[O/H] plot, an upturn in [C/O] is detected towards lower metallicities 
(Fig.~\ref{fig:cn}, panel a, filled triangles). \citet{ake04} corrected the 
oxygen abundances for non-LTE effects, but did not apply any non-LTE correction 
to the [C/O] ratios, under the assumption that non-LTE corrections for the 
permitted \ion{C}{I} and \ion{O}{I} lines are of similar size. Recent non-LTE 
calculations for 43 halo stars by \citet{f09} confirm and strengthen the case 
for an upturn in [C/O] for [O/H]~$< -$2 (Fig.~\ref{fig:cn}, panel a, crosses). 
The sparse data for extremely metal-poor halo giants by 
\citet[][Fig.~\ref{fig:cn}, stars]{s05} should be given a lower weight, because 
unaccounted 3D effects likely affect the carbon molecular lines employed in the 
analysis \citep{asp05}.

\begin{figure}
  \centering
  \includegraphics[width=\columnwidth]{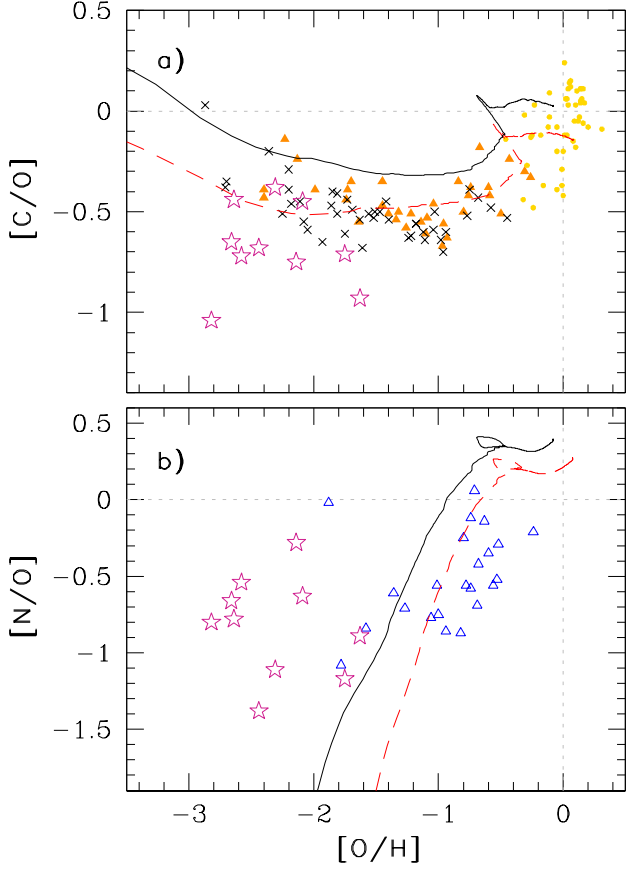}
  \caption{ Carbon-to-oxygen (upper panel) and nitrogen-to-oxygen (lower panel) 
    as functions of [O/H] in the solar neighbourhood. Data (symbols) are the 
    same as in  Fig.~\ref{fig:cn}. Also shown are the predictions from Models~1 
    (short-dashed lines) and 2 (solid lines). The two models differ only in the 
    adopted location of the mass cut in models of SNII explosions. See the 
    electronic edition of the journal for a colour version of this figure.
  }
  \label{fig:ww}
\end{figure}
%

\begin{figure}
  \centering
  \includegraphics[width=\columnwidth]{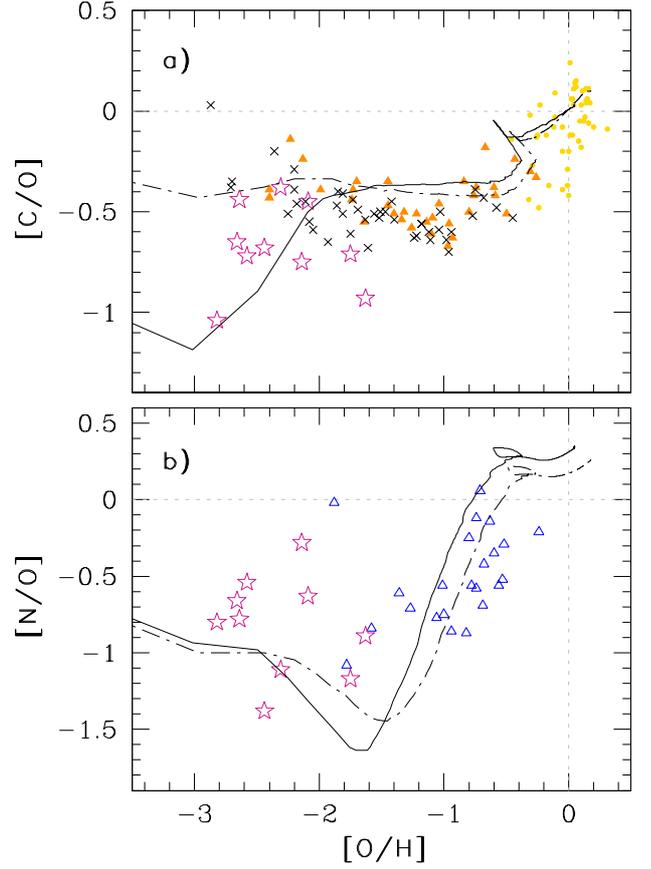}
  \caption{ Same as Fig.~\ref{fig:ww}, but for Models~6 (dot-dashed lines) and 
    7 (solid lines), adopting the pre-SN yields from the Geneva group with and 
    without rotation, respectively. See the electronic edition of the journal 
    for a colour version of this figure.
  }
  \label{fig:rnotr}
\end{figure}

In the [C/O] versus [O/H] diagram (Fig.~\ref{fig:cn}, panel a), the predictions 
from Model~1 [short-dashed (red) line], adopting the metallicity-dependent 
yields of \citet{ww95} case B for massive stars, are in satisfactory agreement 
with the halo data, but fail to reproduce the solar ratio and the trend of 
increasing [C/O] with increasing [O/H] found for the most metal-rich stars. 
\citeauthor{m92}'s \citeyearpar{m92} yields for solar-metallicity stars, 
computed with high mass loss rates but no rotation, tend to overestimate the 
[C/O] ratio at high metallicities [Model~3, long-dashed (black) line]. 
Models~4 and 5, computed using the \citet{k06} yields with 
$\varepsilon_{\mathrm{HN}}$~= 0 and 1, respectively, underestimate the [C/O] 
ratio in the solar vicinity at all metallicities. Since they predict nearly the 
same trend, with differences less than 0.2~dex from one another, we show only 
the results from Model~4 [solid (blue) line]. Models including rotation in 
high-mass stars (i.e. Model~6 and Models from 8 to 15), slightly overestimate 
the [C/O] ratio in the Galactic halo, but are, in general, in good agreement 
with the thin-disc and solar data, apart from Models~10--12, that severely 
overestimate the [C/O] ratios at [O/H]~$> -$0.6. These models use the yields 
from \citet{m01} for LIMSs. To avoid overcrowding, in Fig.~\ref{fig:cn} we show 
the theoretical predictions only for a subset of models, namely Models~6 
[dot-dashed (black) line], 9 [dotted (black) line] and 10 
[short-dashed-long-dashed (blue) line].

In the [C/Fe] versus [Fe/H] plot (Fig.~\ref{fig:cn}, panel c), our models 
predict a decreasing trend of [C/Fe] with increasing metallicity in the 
metallicity domain $-$3.0~$<$ [Fe/H]~$< -$1.0. This trend is followed by an 
increase of [C/Fe] from [Fe/H]~= $-$1.0 to [Fe/H]~= $-$0.5 and, then, by either 
a shallow decline (Models~1, 4, 5, 10) or a flattening of the ratio (Models~3, 
6, 9, 13, 15) towards solar metallicities and up to [Fe/H]~= +0.4. Predictions 
from Models~6, 9, 13 and 15 (the last two not shown in the figure), in 
particular, are in remarkable agreement with data from \citet[][filled circles 
in Fig.~\ref{fig:cn}]{bf06} -- free from non-LTE effects -- and 
\citet[][crosses in Fig.~\ref{fig:cn}]{f09} -- taking appropriate non-LTE 
corrections into account. A notable exception is Model~5, computed with 
\citet{k06} yields for massive stars assuming that 100\% of stars with $m \ge$ 
20~M$_{\sun}$ explode as HNe [$\varepsilon_{\mathrm{HN}}$~= 1, lower (black) 
solid curve in Fig.~\ref{fig:cn}, panel c], which underestimates the [C/Fe] 
ratio across the whole metallicity range.

\begin{figure}
  \centering
  \includegraphics[width=\columnwidth]{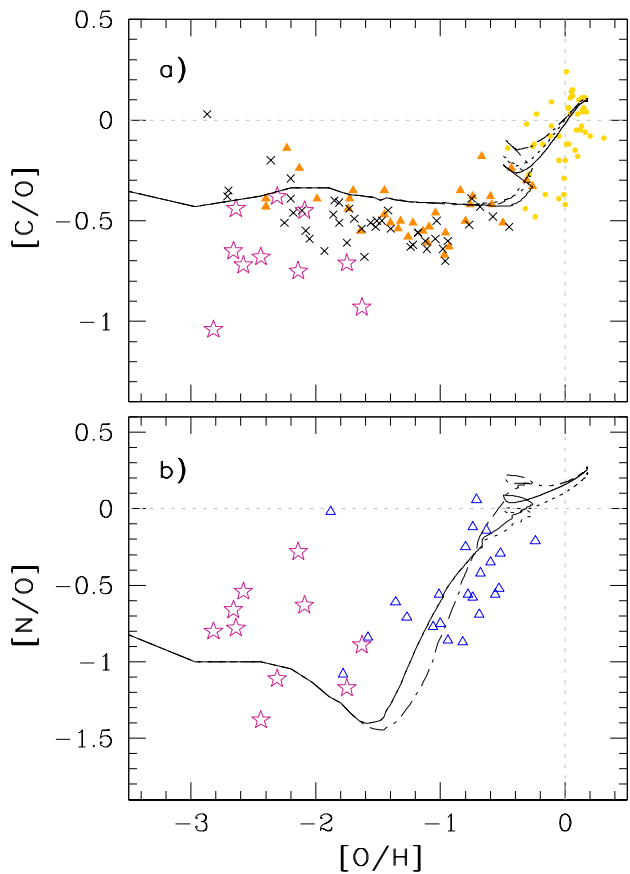}
  \caption{ Same as Fig.~\ref{fig:ww}, but for Models~6 (dot-dashed lines), 8 
    (solid lines) and 9 (dotted lines). Models~6 and 8 differ only in the 
    adopted parameter of mass loss along the AGB, Models~8 and 9 in the assumed 
    mass range in which HBB can take place. See the electronic edition of the 
    journal for a colour version of this figure.
  }
  \label{fig:vdhg}
\end{figure}
%

\begin{figure}
  \centering
  \includegraphics[width=\columnwidth]{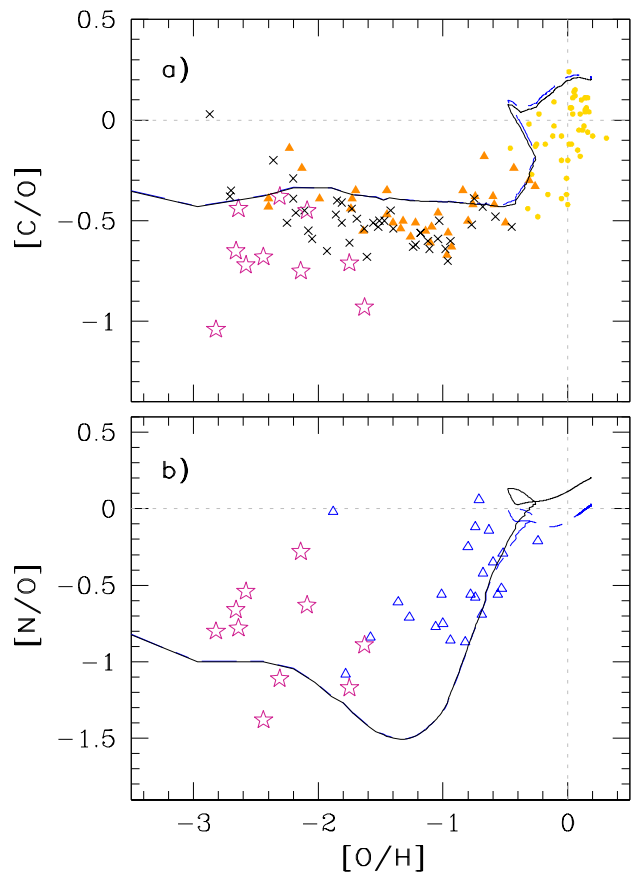}
  \caption{ Same as Fig.~\ref{fig:ww}, but for Models~10 
    (short-dashed-long-dashed lines) and 11 (solid lines), which differ only in 
    the assumed mixing-length parameter for intermediate-mass star models. See 
    the electronic edition of the journal for a colour version of this figure.
  }
  \label{fig:mar}
\end{figure}

All in all, we find differences up to 0.5--0.6~dex in the predicted [C/O] 
ratios and up to 0.8~dex in the predicted [C/Fe] ratios. This quantifies the 
uncertainties in the model predictions due to the uncertainties in the adopted 
stellar yields. The models discussed up to now adopt the best yield sets from 
different authors, which differ quite a lot in the assumed input physics (see 
Sect.~\ref{sec:yields}). To evaluate the effect of changing \emph{only one} 
parameter of stellar evolution on the predictions of GCE models, we run: (i) 
Model~2, which is the same as Model~1 but for the assumed final kinetic energy 
of the ejecta from $m \ge$~30 M$_{\sun}$ stars at infinity \citep[which is 
lower in Model~2 than in Model~1; see][case A versus case B]{ww95}; (ii) 
Model~7, which is the same as Model~6 but for the assumed pre-SN yields of He 
and CNO elements from massive stars (which are now taken from the computations 
of the Geneva group for \emph{non-rotating} stellar models); (iii) Model~8, 
which differs from Model~6 only in the assumed parameter of mass loss along the 
AGB (which is constant with metallicity rather than varying with $Z$ in 
Model~8) and from Model~9 in that HBB is allowed to operate in a smaller mass 
range in Model~9; (iv) Model~11, which is the same as Model~10 but for the 
assumed mixing-length parameter for intermediate-mass stars \citep[$\alpha$~= 
2.50 rather than 1.68 -- this parameter mainly affects the production of $^4$He 
and $^{14}$N due to HBB;][]{m01}.

The predictions of Model~2 versus Model~1, Model~7 versus Model~6, Model~8 
versus Models~6 and 9, and Model~11 versus Model~10 regarding the behaviour of 
[C/O] versus [O/H] in the solar neighbourhood are shown in the top panels of 
Figs.~\ref{fig:ww}, \ref{fig:rnotr}, \ref{fig:vdhg} and \ref{fig:mar}, 
respectively. It is seen that changing the mass cut in models of SNII 
explosions has a profound impact on the predicted [C/O] at all metallicities. 
The inclusion of rotation in models of high-mass stars substantially changes 
the predictions on the [C/O] ratio for [O/H]~$< -$2. Smaller (or even 
negligible) effects are seen at higher metallicity in dependence on the assumed 
mass loss and/or HBB efficiency in LIMSs.

\paragraph{\bf Nitrogen} 
Nitrogen is synthesized through the CNO cycle in the star hydrogen-burning 
layers \citep{c83,a96}. It was recognized long ago that LIMSs are net producers 
of (primary and secondary) nitrogen on a Galactic scale \citep{ep78,rv81}. 
While it was suggested years ago that metal-poor massive stars needed to 
significantly contribute to the production of primary nitrogen \citep[see 
e.g.,][]{m86}, only recently has the physical mechanism been identified as 
rotation \citep{mm02b}. 

In the [N/Fe] versus [Fe/H] diagram, the interpretation of the observations and 
the search for any abundance trends are complicated by the fact that the 
stellar abundances show considerable scatter at a fixed [Fe/H], which is 
largely attributable to the weakness of the usable spectral lines and/or to the 
$T_\mathrm{eff}$ sensitivity of the derived abundances. Indeed, in late-type 
stars one can rely only on weak, high-excitation near-IR \ion{N}{I} lines, also 
subject to non-LTE line formation and 3D effects. In halo stars, only the NH 
and CN molecular lines can be observed, and both have disadvantages. The NH 
band-head falls in a crowded region of the spectra, whereas an analysis of the 
CN lines requires a priori knowledge of carbon abundances. Furthermore, both 
lines are affected by 3D effects, that for NH can result in $-$0.6 and $-$0.9 
abundance corrections for turn-off stars at [Fe/H]~= $-$2 and [Fe/H]~= $-$3, 
respectively, with giants having not been investigated yet \citep[][and 
references therein]{asp05}.

The [N/O] versus [O/H] plot is another powerful diagnostic diagram, but once 
again literature data need corrections owing to non-LTE and 3D effects. 
\citet{i04} estimated robust [N/O] ratios for 31 unevolved metal-poor stars 
using nitrogen and oxygen abundances from near-UV NH and OH lines. Although 
their [N/O] (Fig.~\ref{fig:cn}, panel b, empty triangles) should stay basically 
intact, the [O/H] values would decrease by $\sim$~0.5 dex in 3D \citep{asp05}. 
More severe corrections should affect the [N/O] ratios derived by 
\citet[][Fig.~\ref{fig:cn}, panel b, stars]{s05}: \citet{asp05} estimates that 
they would decrease by $\sim$~0.4 dex, while lower negative corrections of 
$\sim$~0.2 dex should apply to the [O/H] values. This would lead to a slowly 
increasing trend of [N/O] with [O/H] even at the lowest metallicities.

Being aware of all the shortcomings affecting both the data and the models, we 
show the predictions of some of our models for [N/O] versus [O/H] and [N/Fe] 
versus [Fe/H] in Fig.~\ref{fig:cn}, panels b and d, respectively. Other 
predictions from models where the parameters of stellar evolution are changed 
one at a time, are shown in Figs.~\ref{fig:ww} to \ref{fig:mar} (only for [N/O] 
versus [O/H]). Model predictions are compared to data from \citet{s05}, 
\citet{l08}, \citet{i04}, and \citet{r03}. It is apparent that only Models~6 to 
15, adopting the N yields from low-metallicity massive stars from the Geneva 
group, are able to explain the current measurements of [N/O] and [N/Fe] in very 
metal-poor halo stars (see also \citealt{e08}, their figure~9). The most 
massive zero-metallicity stars produce great amounts of N independently of the 
assumed stellar rotation (see \citealt{e08}, their table~4). As a consequence, 
a plateau in [N/O], [N/O]~$\sim -$0.9, is found for [O/H]~$< -$2.5 when 
assuming the Geneva group yields (Fig.~\ref{fig:rnotr}, lower panel), 
independently of whether zero-metallicity stars rotate fast or do not rotate at 
all. However, all the models severely underestimate the [N/O] ([N/Fe]) ratio in 
the metallicity range $-$2~$<$ [O/H]~$< -$1 ($-$3~$<$ [Fe/H]~$< -$1). A 
contribution to the N enrichment from the winds of super-AGB stars could 
improve the model predictions. On the other hand, in the framework of our 
models, a range in [Fe/H] from $-$3 to $-$1 corresponds to a range in overall 
metallicity $Z$ from about 4$\times$10$^{-5}$ to about 0.002. In this 
metallicity range, we are using the yields obtained by interpolating the grids 
computed by \citet{mm02a} for massive slow rotators with $Z$~= 10$^{-5}$ and 
$Z$~= 0.004. Thus, the predicted deficiency of nitrogen could unravel the need 
for an upward revision of \citeauthor{mm02a}'s \citeyearpar{mm02a} N yields. 
This could be achieved by making the stars rotate faster at those metallicities 
\citep[the faster the stellar rotation, the larger the primary N 
production;][]{mm02b}.

Changing the prescriptions about N production in LIMSs has a profound impact on 
the theoretical predictions at higher metallicities, [O/H]~$> -$1.4 and 
[Fe/H]~$> -$2.0. It is clear that both changing the rate of mass loss along the 
AGB and the efficiency of HBB in models of intermediate-mass stars play a 
significant role in shaping the model predictions for N (see the lower panels 
of Figs.~\ref{fig:vdhg} and \ref{fig:mar}, respectively). Differences up to 
0.7~dex in both [N/O] and [N/Fe] are found at intermediate metallicities for 
models keeping the same nucleosynthesis prescriptions for massive stars and 
changing those for LIMSs (cfr. predictions from Models~9 and 10, dotted and 
short-dashed-long-dashed lines, respectively, in Fig.~\ref{fig:cn}, panels~b 
and d). Model~15, which assumes the revised N yields by \citet{k10} for LIMSs, 
is the only one which fits best both the solar [N/O] and [N/Fe] ratios 
(Fig.~\ref{fig:cn}, thick solid line).

\paragraph{\bf Oxygen} 
\begin{figure}
  \centering
  \includegraphics[width=\columnwidth]{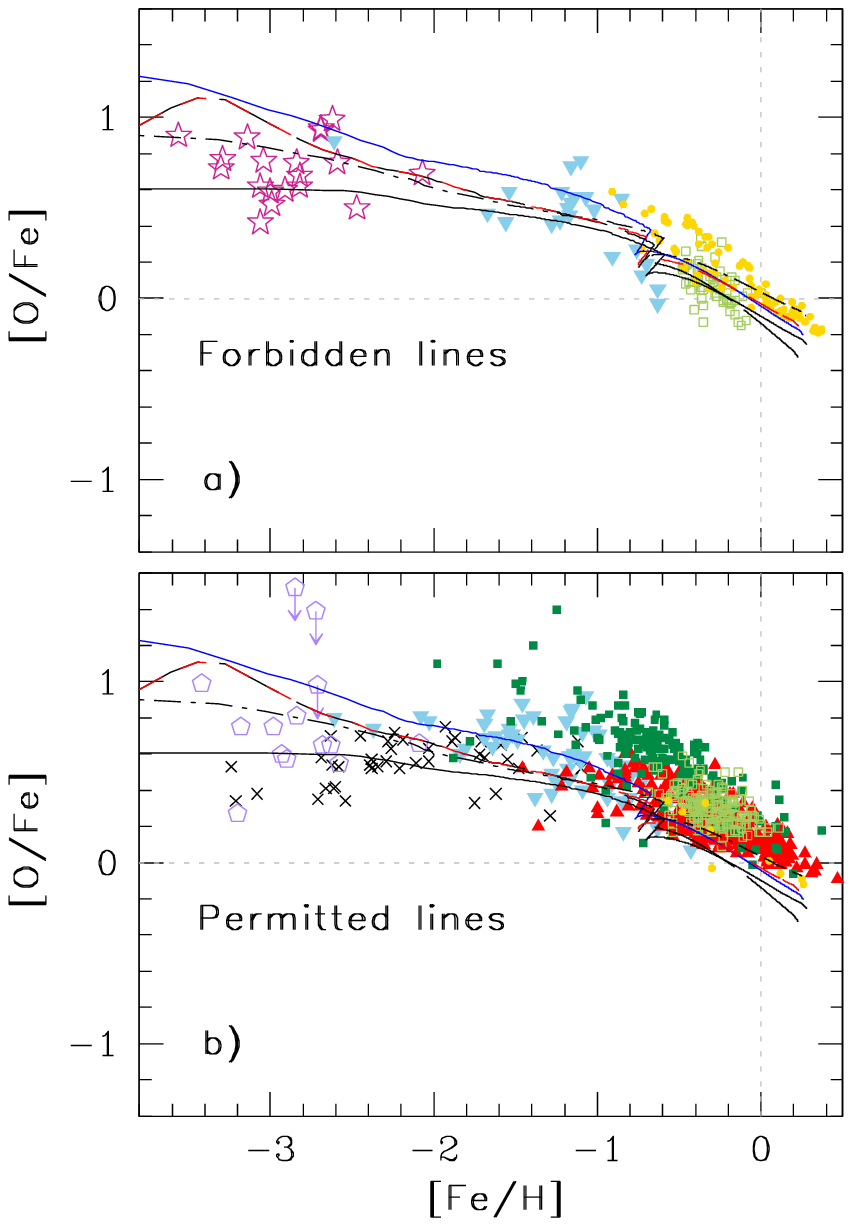}
  \caption{ [O/Fe] versus [Fe/H] in the solar neighbourhood. Theoretical 
    predictions are from Models~1 [short-dashed (red) lines], 3 [long-dashed 
    (black) lines], 4 [upper (blue) solid lines], 5 [lower (black) solid lines] 
    and 6 [dot-dashed (black) lines]. Upper panel: oxygen abundances from the 
    forbidden [\ion{O}{I}] lines from \citet[][stars]{c04}, 
    \citet[][upside-down triangles]{g03}, \citet[][open squares]{r03} and 
    \citet[][filled circles]{b05}. Lower panel: oxygen abundances from other 
    indicators from \citet[][open pentagons]{l08}, \citet[][crosses]{f09}, 
    \citet[][upside-down triangles]{g03}, \citet[][filled squares]{rlap06}, 
    \citet[][open squares]{r03}, \citet[][triangles]{rapl07} and 
    \citet[][filled circles]{b05}. See the electronic edition of the journal 
    for a colour version of this figure.}
  \label{fig:o}
\end{figure}
%
\begin{figure}
  \centering
  \includegraphics[width=\columnwidth]{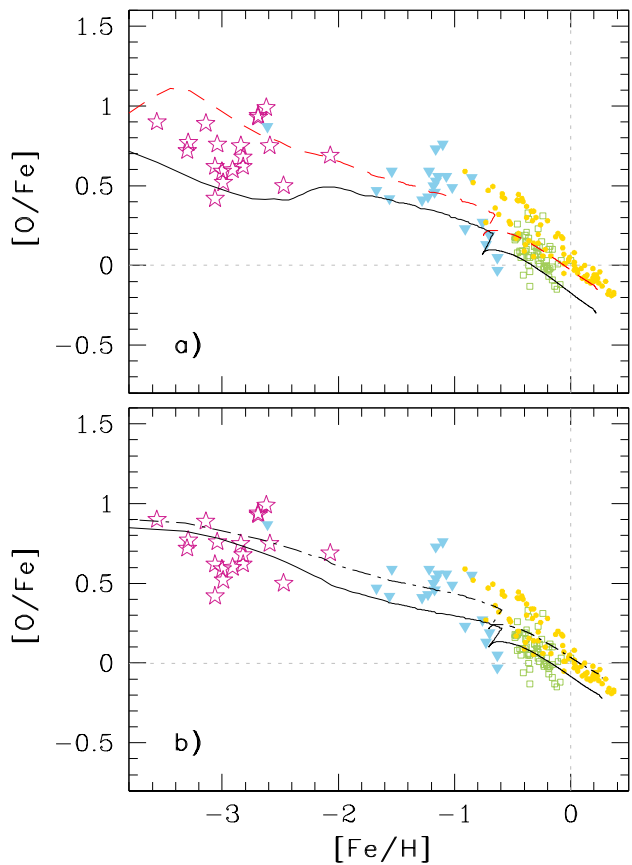}
  \caption{ Upper panel: predictions on [O/Fe] versus [Fe/H] from Models~1 and 
    2 [short-dashed (red) and solid (black) lines, respectively] are compared 
    one to another and with data from the forbidden [\ion{O}{I}] lines (see 
    Fig.~\ref{fig:o} caption for references). Lower panel: same as the upper 
    panel, but for Models~6 and 7 [dot-dashed and solid (black) lines, 
    respectively]. See the electronic edition of the journal for a colour 
    version of this figure.}
  \label{fig:obis}
\end{figure}
Next to hydrogen and helium, oxygen is the third most abundant element in the 
universe. It forms essentially in the interiors of massive stars, through 
hydrostatic burning of He, C, and Ne \citep[e.g.][]{a96}. Available indicators 
of oxygen abundances in stars include the \ion{O}{I} triplet at $\sim$7774~\AA 
\ (mainly observed in FG dwarfs and subgiants), the forbidden [\ion{O}{I}] 
lines at 6300 and 6363~\AA \ (detected in giants and cool subgiants, and barely 
detectable in dwarfs), and the OH lines in the UV (observed in FGK metal-poor 
stars) and in the IR (only observed in cool stars with $T_\mathrm{eff} <$ 
5000~K). The features at 7774~\AA \ are strong and located in a clean part of 
the spectrum but subject to large and poorly understood deviations from LTE 
conditions. The forbidden lines are free from non-LTE effects but weak and 
blended by lines from other chemical species. Furthermore, switching from 1D to 
3D models changes the [\ion{O}{I}]-based abundances by as much as $-$0.2 dex in 
turn-off stars with [Fe/H]~= $-$2; a similar behaviour can be anticipated for 
metal-poor giants \citep{n02}. The OH lines are difficult to observe and, in 
the UV domain, are also affected by surface inhomogeneities. The most reliable 
indicators of oxygen abundances in stars seem to be the forbidden [\ion{O}{I}] 
lines (see volume~45 issue~8 of New Astronomy), especially if spectra of high 
resolution and high S/N are acquired and analyzed using accurate atomic line 
data \citep{n02}.

In Fig.~\ref{fig:o} we compare some of our oxygen model predictions to 
observations of the forbidden [\ion{O}{I}] line as well as of other oxygen 
lines (panels a and b, respectively), concentrating on models which adopt the 
best yield sets suggested by different authors. Models~1, 6, 8, 9, 10, 11, 13 
and 15\footnote{For sake of clarity, only the predictions from Models~1 and 6 
are shown in Fig.~\ref{fig:o} [short-dashed (red) and dot-dashed (black) lines, 
respectively].} forecast an [O/Fe] ratio which, starting from [O/Fe]~$\simeq 
+$1 at [Fe/H]~$\simeq -$4, slopes gently down to subsolar values at the highest 
metallicities. This is in excellent agreement with the observations, especially 
if the [\ion{O}{I}]-based abundances are considered for comparison. The gradual 
increase of [O/Fe] with decreasing [Fe/H] in the [Fe/H]~$< -$1 regime stems 
from the larger oxygen-to-iron ratios in the ejecta of more massive, more 
metal-poor SNeII. The quick decrease of [O/Fe] above [Fe/H]~$\simeq -$1 is due 
instead to the overwhelming contribution to the Fe production by SNeIa. 
Models~3, 12 and 14, assuming \citeauthor{m92}'s \citeyearpar{m92} yields for 
solar-metallicity massive stars, predict a decrease of the [O/Fe] ratio at high 
metallicities definitely steeper than indicated by trustworthy data from 
\citet[][Fig.~\ref{fig:o}, panel a, filled circles]{b05}. In order to avoid 
overcrowding, only the predictions from Model~3 are shown [Fig.~\ref{fig:o}, 
long-dashed (black) curves]. Model~4, adopting the yields by \citet{k06} for 
normal SNe, fits the disc and Sun data nicely [Fig.~\ref{fig:o}, upper (blue) 
solid curve]. However, it slightly overestimates the halo data. Model~5, 
instead, assuming that all stars above 20~M$_{\sun}$ explode as HNe with much 
higher explosion energies, predicts an almost flat [O/Fe] versus [Fe/H] trend 
in the early halo [Fig.~\ref{fig:o}, lower (black) solid line]. Though most of 
the observed [O/Fe] ratios lie above the theoretical predictions, there are 
some stars whose abundances are consistent with a scenario of formation from 
pure HN ejecta. 

It is common belief that the synthesis of O in stars is well understood. Yet, 
by using different prescriptions on the O production in stars, we find that the 
predicted [O/Fe] ratio may vary by as much as 0.6~dex at the lowest 
metallicities. The difference reduces to 0.2~dex at solar metallicity. At the 
origin of the uncertainties in the predictions of our GCE model there is, once 
again, the complex interplay among the different assumptions made by different 
authors working on stellar evolution and nucleosynthesis to treat the different 
processes regulating the evolution of the stars. In Fig.~\ref{fig:obis}, upper 
panel, we compare the predictions of Models~1 and 2, which differ only in the 
location of the mass cut in models of SNII explosions. It is seen that a 
variation in this parameter may have a huge impact (up to 0.5~dex) on the 
predicted [O/Fe] at the lowest metallicities. Stellar rotation has a lower -- 
still non-negligible -- impact (see Fig.~\ref{fig:obis}, lower panel).

\subsection{Odd-Z elements}
\label{sec:oddz}

\paragraph{\bf Sodium} 
Sodium is an odd-Z element, synthesized mostly during hydrostatic carbon 
burning, and partly during hydrogen burning through the NeNa cycle; the 
$s$-process produces some Na as well \citep{c03}. Adopting the LTE 
approximation in standard abundance analyses of Na can lead to substantial 
overestimates of the Na abundances. Non-LTE abundance corrections can be very 
large, typically $-$0.1~dex at solar metallicity and $-$0.5 dex or more at 
[Fe/H] around $-$2, for the 589 and 819 nm doublets; the departures from LTE 
become less severe for [Fe/H]~$< -$2 as the line formation shifts to deeper 
layers \citep[e.g.][]{bbg98}. The weaker 568 and 615 nm doublets are less 
sensitive to non-LTE effects. Departures from LTE are most likely the culprits 
for the discordant values of [Na/Fe] reported in the literature for metal-poor 
stars \citep[see][and references therein]{rlap06}. From their LTE analysis of 
the \ion{Na}{I} lines at 6154 and 6160 \AA, \citet{rlap06} find a hint for an 
increase of [Na/Fe] with increasing [Fe/H] in the metallicity range $-$1.0~$<$ 
[Fe/H]~$< -$0.6, followed by a decrease towards solar values for $-$0.6~$<$ 
[Fe/H]~$<$ 0. Such a trend is confirmed by \citet{sgz04} using non-LTE 
statistical equilibrium calculations. For [Fe/H]~$>$~0, \citet{bfl03,b05} 
observe a rise in the [Na/Fe] trend, which was also seen by \citet{e93}. As for 
the lowest metallicities, when only extremely metal-poor dwarfs and unmixed 
giants are considered and the non-LTE corrections of the Na abundances are 
taken into account, one gets a flat trend of the [Na/Fe] ratio in the 
metallicity regime $-$4.0~$<$ [Fe/H]~$< -$2.5, [Na/Fe]$_{\mathrm{NLTE}}$~= 
$-$0.21~$\pm$ 0.13 dex, and a small star-to-star variation, comparable to that 
found for other elements \citep{a07}.

\begin{figure}[b!]
  \centering
  \includegraphics[width=\columnwidth]{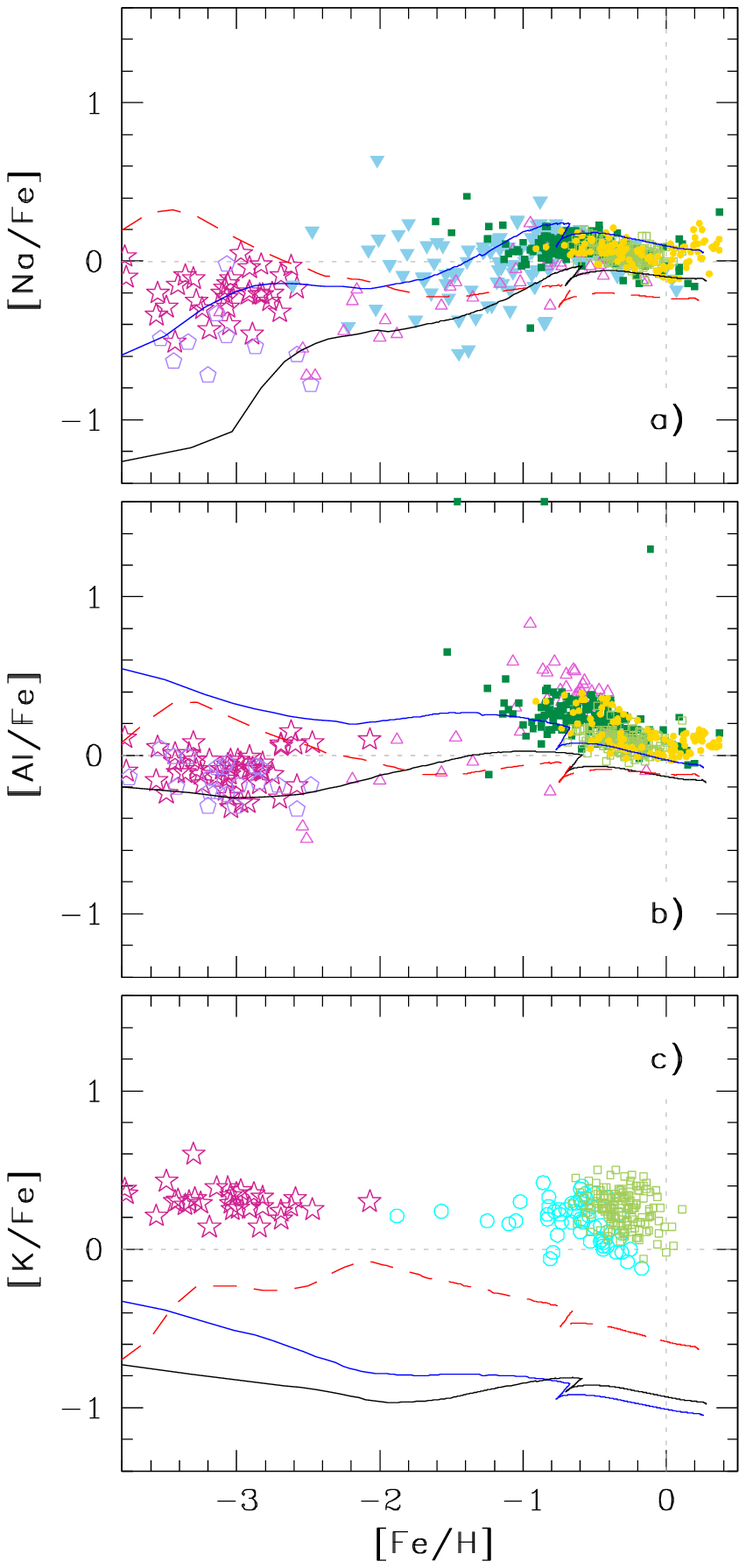}
  \caption{ [X/Fe] versus [Fe/H] relations for sodium (panel a), aluminium 
    (panel b) and potassium (panel c) in the solar neighbourhood. Theoretical 
    predictions are from Models~1 [short-dashed (red) lines], 4 [upper solid 
    (blue) lines] and 5 [lower solid (black) lines]. Data are from 
    \citet[][stars, for Na, Al and K, respectively]{a07,a08,a10}, \citet[][open 
    pentagons]{l08}, \citet[][upside-down triangles]{g03}, \citet[][open 
    triangles]{g06}, \citet[][open circles]{z06}, \citet[][filled 
    squares]{rlap06}, \citet[][open squares]{r03} and \citet[][filled 
    circles]{b05}. LTE abundances from \citet{l08} have been corrected by 
    $-$0.5 dex for Na \citep{bbg98} and +0.65 dex for Al \citep{bg97}. See the 
    electronic edition of the journal for a colour version of this figure.}
  \label{fig:naalk}
\end{figure}

As one can see in Fig.~\ref{fig:naalk}, panel a, only Model~4 [upper solid 
(blue) curve], corresponding to the yields by \citet{k06} for massive stars 
without HNe, predicts a [Na/Fe] ratio in agreement with the halo data. Models 
including nucleosynthesis from HNe, namely, Models from 5 to 15 [for sake of 
clarity, only the predictions from Model~5 are shown in the figure, lower solid 
(black) curve], underestimate the [Na/Fe] ratio in the halo. Model~1 
[short-dashed (red) line], adopting the yields of \citet{ww95} case B, provides 
an acceptable fit to the halo ratios, but only in the $-$2.5~$\le$ [Fe/H]~$\le 
-$1 metallicity range. All in all, at the lowest metallicities we find 
differences up to 1.4~dex in the predicted [Na/Fe] ratio which, similarly to 
what we found for the [O/Fe] ratio, are largely driven by the adopted mass cut 
and energy of the explosion in models of massive stars. These differences 
reduce to $\sim$0.2~dex at [Fe/H]~$> -$2.

\begin{figure}
  \centering
  \includegraphics[width=\columnwidth]{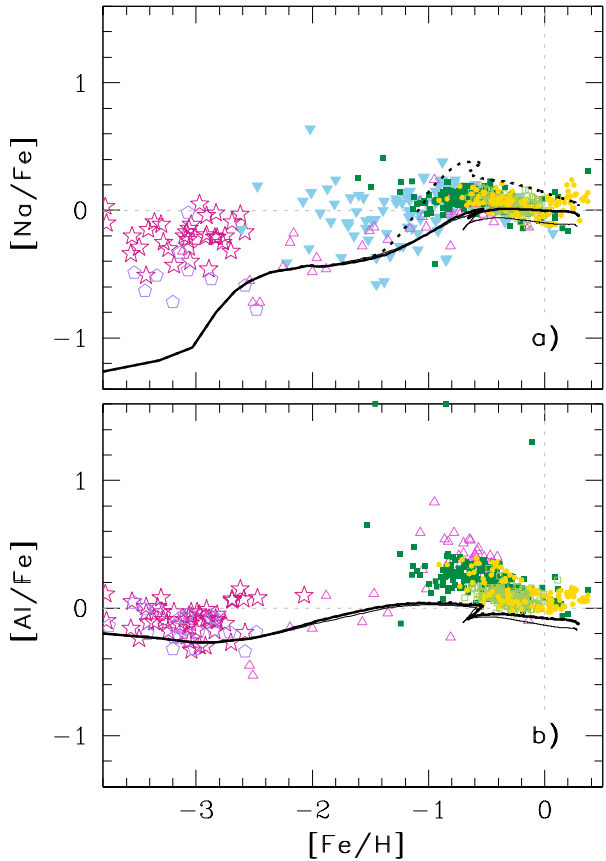}
  \caption{ Same as Fig.~\ref{fig:naalk}, panels~a and b, but for Models~5 
    (thin solid lines), 13 (thick dotted lines) and 15 (thick solid lines). The 
    predictions for [Al/Fe] versus [Fe/H] from Model~15 lay on those from 
    Model~13. See the electronic edition of the journal for a colour version of 
    this figure.}
  \label{fig:naalfe}
\end{figure}

Models from 1 to 12 do not account for Na synthesis from LIMSs. If Na 
production from LIMSs is considered, very different predictions for Na 
evolution in the disc are found, depending on the adopted yields. Model~15, 
adopting the revised \citet{k10} yields of Na from LIMSs 
(Fig.~\ref{fig:naalfe}, panel a, thick solid line) predicts an almost 
negligible production of sodium from these stellar sources on a Galactic scale. 
The adoption of the old \citet{kl07} yields, instead, leads to a huge Na 
production from LIMSs in the disc (Fig.~\ref{fig:naalfe}, panel a, thick dotted 
line). These differences in model predictions for Na are essentially due to the 
up-to-date nuclear reaction rates adopted by \citet[][see discussion in 
Sect.~\ref{sec:yields}]{k10}. None of the models, however, is able to explain 
the intriguing upward trend of [Na/Fe] at [Fe/H]~$>$ 0, first pointed out by 
\citet{e93} and clearly seen in the \citet{b05} data (filled circles in 
Fig.~\ref{fig:naalfe}). None of the models includes Na production from rotating 
massive stars. We will come back to this point in Sect.~\ref{sec:conc}.

%

\paragraph{\bf Aluminium} 
Aluminium is a product of hydrostatic carbon and neon burning 
\citep[e.g.][]{ww95}. It can also be produced during hydrogen burning in the 
MgAl chain. Observationally, Al abundances need large non-LTE corrections: at 
solar [Fe/H], they amount to approximately +0.1 dex for the \ion{Al}{I} 396~nm 
resonance line, but reach +0.4 dex or even more for [Fe/H]~$< -$1.0; the 
maximum correction, +0.8 dex, is required for mildly metal-poor turn-off stars 
in which the line is saturated \citep{asp05}. For [Fe/H]~$> -$1.0, Al shows a 
`Mg-like behaviour' \citep{bfl03,b05,rlap06}. At very low metallicities, 
[Fe/H]~$< -$2.8, the relation [Al/Fe] versus [Fe/H] becomes rather flat, 
[Al/Fe]$_{\mathrm{NLTE}}$~= $-$0.06~$\pm$ 0.10 \citep{a08}; the scatter around 
the mean value is smaller than the scatter of the LTE determinations 
\citep{c04} and can be explained by measurement errors only. A few sparse data 
are available in the intermediate metallicity regime. Here, the non-LTE [Al/Mg] 
ratio displays an intriguing step-like difference between stars belonging to 
the Galactic halo and disc populations, that might prove useful as a further 
membership diagnostic criterion \citep[see][and discussion therein]{g06}. We 
will come back to this issue later.

In the low-metallicity domain, the best-fitting models are those adopting 
\citet{k06} yields for massive stars with $\varepsilon_{\mathrm{HN}}$~= 1 (see 
Table~\ref{tab:nucp}; only predictions from Model~5 are shown in 
Fig.~\ref{fig:naalk}). Such models predict an almost constant [Al/Fe] ratio 
during the first 370 Myr of Galactic evolution. Then, the ratio increases 
mildly, owing to the metal dependency of Al production in massive stars, and 
flattens again around a Galactic age of 1~Gyr, because of the increasing Fe 
production from SNeIa. None of the models can explain the behaviour of [Al/Fe] 
in the Galactic disc. The activation of the MgAl chain in the central regions 
of rotating massive stars on the main sequence, followed by transportation of 
Al-rich matter to the outer envelope and ejection of the outermost layers by 
slow stellar winds, might lead to a substantial modification of the Al yields 
used here \citep{d07}. However, complete grids of Al yields from rotating 
massive stars are not available yet. Also, one must be aware that, in order to 
reproduce the Mg-Al anticorrelation observed in globular clusters stars, 
\citet{d07} varied the rate of the nuclear reaction $^{24}$Mg(p, 
$\gamma$)$^{25}$Al by orders of magnitude outside of the published 
uncertainties.

The role of Al production from LIMSs should be better investigated as well: the 
recent study by \citet{k10} ascribes almost no Al production to LIMSs 
(Fig.~\ref{fig:naalfe}, lower panel, thick solid line), confirming results from 
previous work by \citet[][Fig.~\ref{fig:naalfe}, lower panel, thick dotted 
line]{kl07}. If AGB stars are responsible for the globular cluster Mg-Al 
anticorrelation, then it is likely that AGB stars as a whole synthesize more Al 
than predicted by current models \citep{mar09}. Should a significant amount of 
Al come from LIMSs, a sudden rise of the [Al/Fe] ratio in field Galactic stars 
would be obtained starting at [Fe/H]~$\sim -$1.5 (cfr. Fig.~\ref{fig:naalfe}, 
panel a, thick dotted line, showing the results of Model~13, with a great Na 
production from LIMSs), in agreement with the trend inferred from the 
observations.

Although the uncertainties in the GCE model predictions for Al are by far less 
dramatic than those for Na, especially in the high metallicity domain, the 
comparison with the available Al abundance data clearly shows that the 
synthesis of Al at disc metallicities is still far from having been understood.
 
\paragraph{\bf Potassium} 
The alkali element potassium is mostly produced by a combination of hydrostatic 
oxygen shell burning and explosive oxygen burning in proportions that vary 
depending on the stellar mass \citep[e.g.][and references therein]{ww95}. 
LTE abundances obtained from the \ion{K}{I} 769.9~nm resonance line are 
overestimated by 0.4--0.7 dex over a wide range of metallicities; non-LTE 
corrections are smaller, but remain significant (from $-$0.15 dex to $-$0.3 
dex), for the $\lambda\lambda$~12522, 12432, and 11769~\AA \ weak subordinate 
lines \citep{is00}.

In Fig.~\ref{fig:naalk}, panel c, we compare our potassium model predictions to 
the data by \citet[][]{a10}, \citet[][]{z06} and \citet[][]{r03}. Non-LTE 
corrections are taken into account in the first two studies. Variations up to 
1~dex in [K/Fe] are found among different model predictions. The comparison 
between data and model predictions shows the absolute inadequacy of the adopted 
stellar yields.

\subsection{Alpha elements}
\label{sec:alpha}

\begin{figure}
  \centering
  \includegraphics[width=\columnwidth]{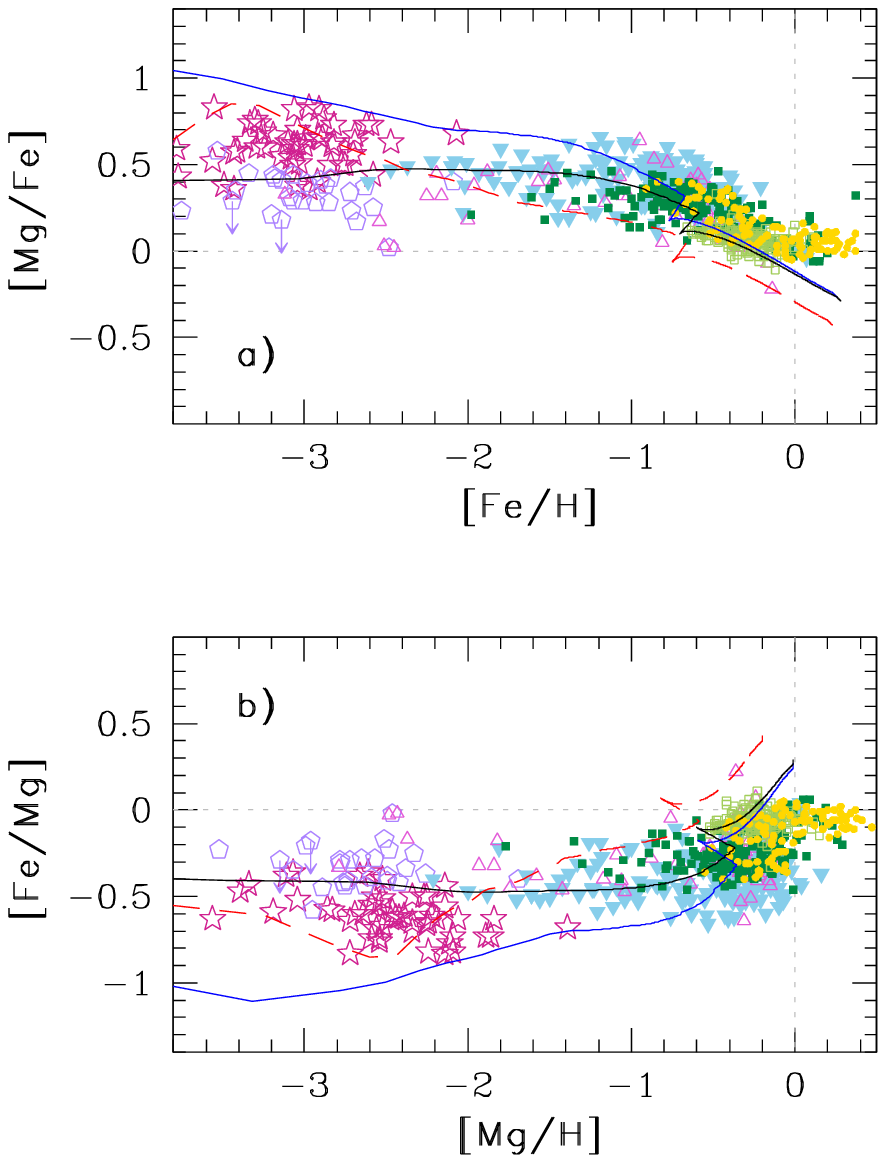}
  \caption{ [Mg/Fe] versus [Fe/H] (top panel) and [Fe/Mg] versus [Mg/H] (bottom 
    panel) in the solar neighbourhood. Theoretical predictions are from 
    Models~1 [short-dashed (red) lines], 4 [upper solid (blue) line in panel a 
    and lower solid (blue) line in panel b] and 5 [lower solid (black) line in 
    panel a and upper solid (black) line in panel b]. Data are from 
    \citet[][stars]{a10}, \citet[][open pentagons]{l08}, \citet[][upside-down 
    triangles]{g03}, \citet[][open triangles]{g06}, \citet[][filled 
    squares]{rlap06}, \citet[][open squares]{r03} and \citet[][filled 
    circles]{b05}.  See the electronic edition of the journal for a colour 
    version of this figure.}
  \label{fig:mg}
\end{figure}

\paragraph{\bf Magnesium} 
Most of magnesium comes from hydrostatic carbon burning and explosive neon 
burning in massive stars \citep[e.g.][]{ww95}. Stellar spectroscopists may rely 
on several \ion{Mg}{I} lines, as well as on a few \ion{Mg}{II} lines, to derive 
magnesium abundances. Non-LTE corrections are of little concern in this case; 
they are negative for the high-excitation \ion{Mg}{II} lines and the 
\ion{Mg}{I}~b triplet at 517~nm ($-$0.05 dex and $-$0.2 dex at most at low 
[Fe/H], respectively), while they are positive for the remaining lines, the 
maximum correction being +0.2 dex at low [Fe/H] \citep[][and references 
therein; but see \citealt{a10}, pointing to somehow larger corrections]{asp05}.

\begin{figure}[b!]
  \centering
  \includegraphics[width=\columnwidth]{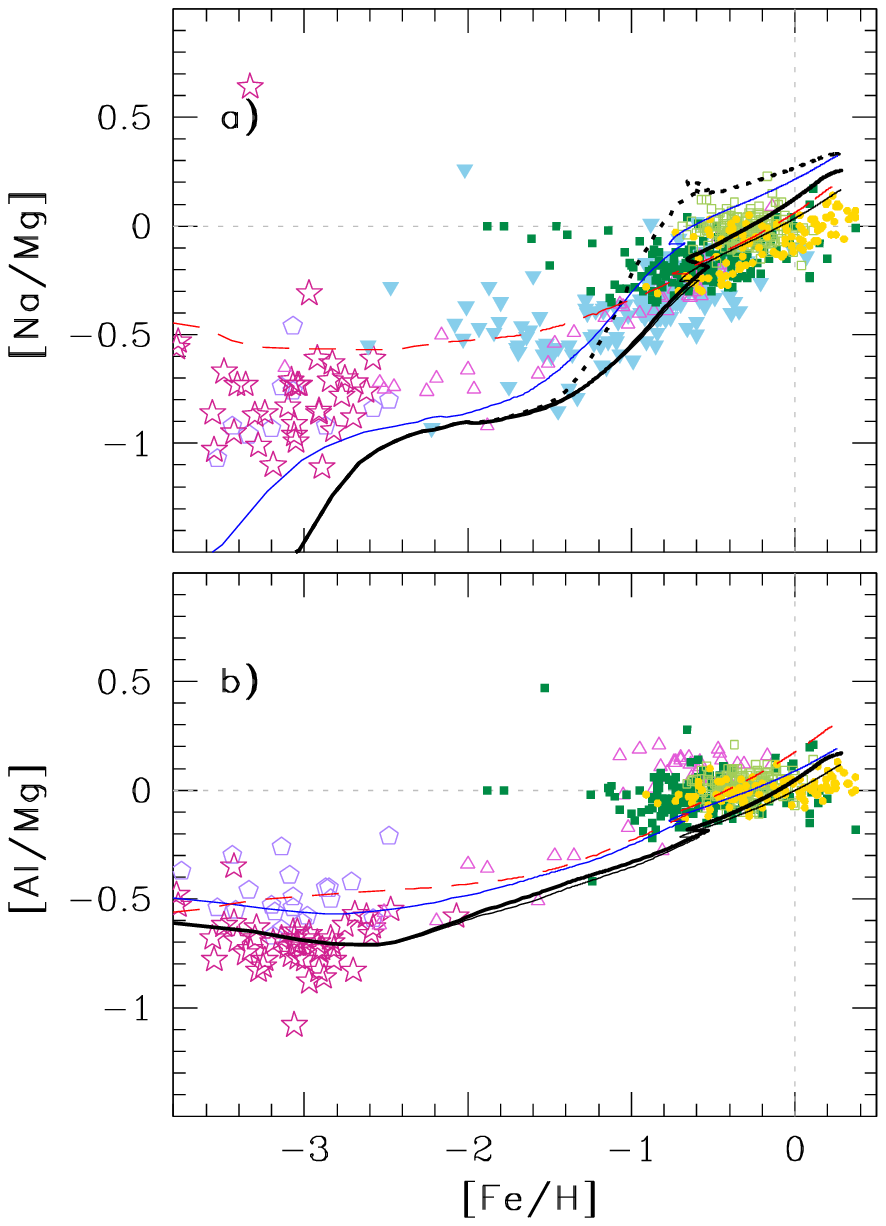}
  \caption{ [X/Mg] versus [Fe/H] relations for sodium (panel a) and aluminium 
    (panel b) in the solar neighbourhood. Theoretical predictions are from 
    Models~1 [short-dashed (red) lines], 4 [upper solid (blue) lines], 5 [lower 
    solid (black) lines], 13 (thick dotted line) and 15 (thick solid lines). 
    Data are from \citet[][stars, for Na, Al and Mg abundances, 
    respectively]{a07,a08,a10}, \citet[][open pentagons]{l08}, 
    \citet[][upside-down triangles]{g03}, \citet[][open triangles]{g06}, 
    \citet[][filled squares]{rlap06}, \citet[][open squares]{r03} and 
    \citet[][filled circles]{b05}. LTE abundances from \citet{l08} have been 
    corrected by $-$0.5 dex for Na \citep{bbg98} and +0.65 dex for Al 
    \citep{bg97}. See the electronic edition of the journal for a colour 
    version of this figure.}
  \label{fig:naalmg}
\end{figure}

GCE models adopting the yields of Mg by \citet{ww95} for massive stars 
underestimate the abundance of magnesium in the discs of the Galaxy \citep[see 
Fig.~\ref{fig:mg}, short-dashed lines, referring to case~B; see 
also][]{t95,gp00}, unless ad hoc adjustments are made to the yields 
\citep{f04}. While detailed nucleosynthesis calculations still have to 
establish whether the modifications proposed by \citet{f04} have a physical 
ground, HNe can provide an alternate solution: \citet{k06} have obtained a nice 
fit to the [Mg/Fe] versus [Fe/H] relation traced by solar neighbourhood stars 
using a GCE model for the solar vicinity which adopts their new nucleosynthetic 
yields for SNeII and HNe, $\varepsilon_{\mathrm{HN}}$~= 0.5 for the HN fraction 
and a Salpeter IMF. Yet, their model can not reproduce values of [Mg/H] 
exceeding solar, at odds with observations \citep[see][their figure~9]{k06}. 
Here we reasses the problem of Mg evolution, adopting a different GCE model, 
with a steeper IMF \citep{k93}, better suited to describe the stellar 
populations of the Milky Way field \citepalias[see][and references 
therein]{r05}, and slightly different mass limits for the IMF, $m_l$~= 
0.1~M$_{\sun}$ and $m_u$~= 100~M$_{\sun}$, rather than $m_l$~= 0.07 M$_{\sun}$ 
and $m_u$~= 50 M$_{\sun}$ as in \citet{k06}. If we trust the non-LTE Mg 
abundances recently computed by \citet{a10} for the most metal-poor stars, the 
best fit to the observed [Mg/Fe] versus [Fe/H] and [Fe/Mg] versus [Mg/H] 
relations is obtained with \citet{k06} yields and $\varepsilon_{\mathrm{HN}}$ 
values intermediate between the two extreme cases analysed here (solid lines in 
Fig.~\ref{fig:mg}). On the other hand, if the LTE Mg abundances by \citet{l08} 
are the ones to be trusted, the extreme case of $\varepsilon_{\mathrm{HN}}$~= 1 
is favoured instead. None of the models, however, is able to reproduce the 
highest [Mg/H] values and the flattening of the relations at nearly 
solar/higher than solar metallicities. This could suggest the need for either a 
revision of current SNII and/or HN yields for solar and/or higher than solar 
metallicity stars, or larger contributions to Mg production from SNeIa, or 
significant Mg synthesis in LIMSs, or a combination of all these factors. 
Clearly, it is not safe to use Mg as the prime metallicity indicator in 
metal-rich galaxies.

\begin{figure*}
  \centering
  \includegraphics[width=\textwidth]{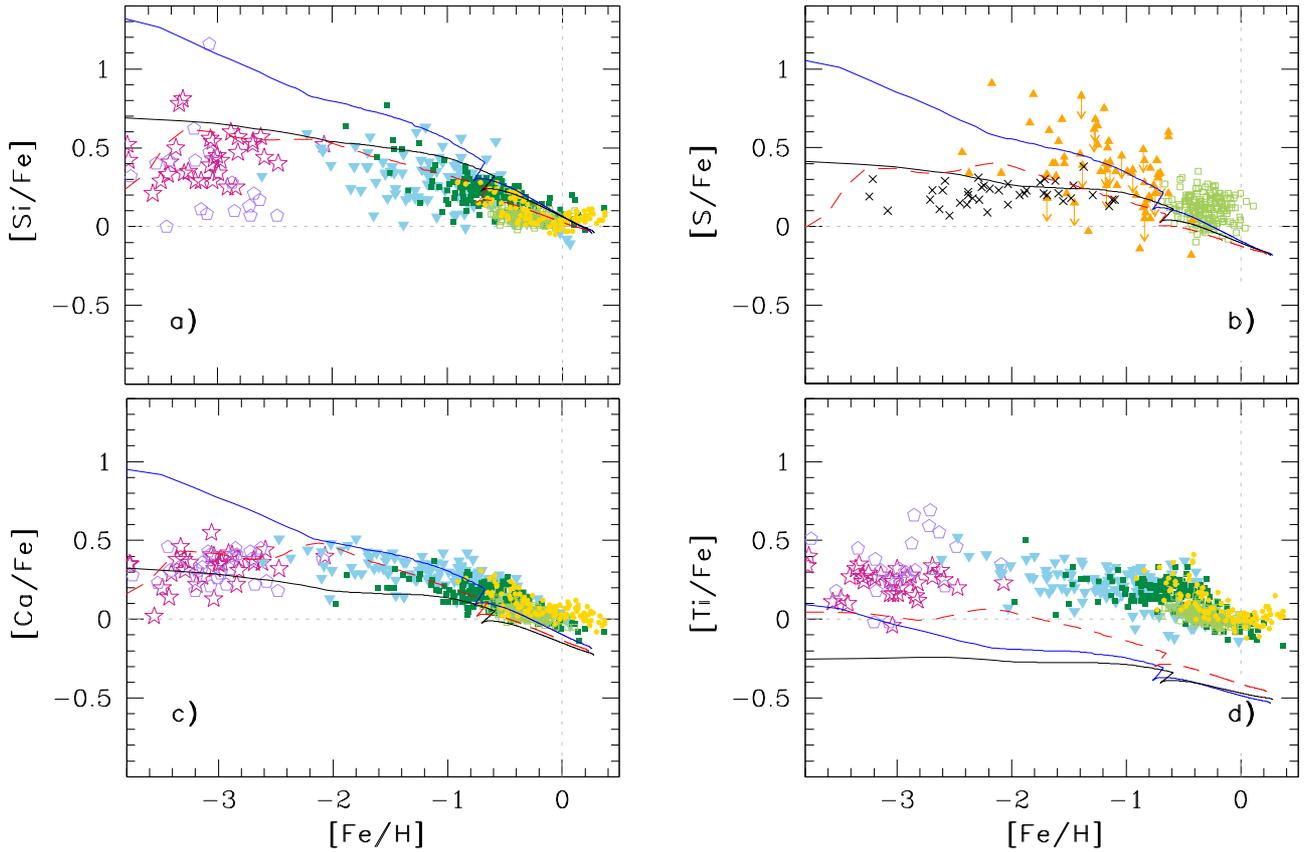}
  \caption{Predicted behaviour of the relative ratios of several $\alpha$ 
    elements to iron (panel a: silicon, panel b: sulphur, panel c: calcium, 
    panel d: titanium) with respect to the relative iron abundance in the solar 
    neighborhood, for Models~1 [short-dashed (red) lines], 4 [upper solid 
    (blue) lines] and 5 [lower solid (black) lines]. Also shown are data from 
    \citet[][stars]{c04}, \citet[][open pentagons]{l08}, 
    \citet[][crosses]{n07}, \citet[][upside-down triangles]{g03}, 
    \citet[][triangles]{c05}, \citet[][filled squares]{rlap06}, \citet[][{open 
      squares}]{r03} and \citet[][filled circles]{b05}. See the electronic 
    edition of the journal for a colour version of this figure.}
  \label{fig:alpha}
\end{figure*}

It has been recently pointed out by \citet{g06} that, while the trend of 
[Na/Mg] versus [Fe/H] is witnessing a history of continuous enrichment from the 
halo to the discs, the [Al/Mg] ratio displays a much more interesting step-like 
behaviour, with halo stars sharing a common value of $\langle$[Al/Mg]$\rangle 
\sim -$0.45 and the two disc populations mostly consistent with 
$\langle$[Al/Mg]$\rangle \sim +$0.10. At the lowest metallicities, say $-$3.6 
$<$ [Fe/H]~$< -$2.5, extremely metal-poor stars share a lower value, 
$\langle$[Al/Mg]$\rangle \sim -$0.7 \citep{a10}. In Fig.~\ref{fig:naalmg} we 
show these data, together with measurements from other authors. Superimposed 
are our theoretical tracks, for a significant subset of models (lines). Three 
results are immediately apparent from this figure. The first is that all models 
overpredict the [Na/Mg] and [Al/Mg] ratios at [Fe/H]~$>$ 0, thus reinforcing 
the demand for a Mg production higher than predicted by current GCE models at 
late times. The second important result is that the step-like behaviour of 
[Al/Mg] versus [Fe/He], suggested by \citet{g06}, can not be reproduced by the 
models. A jump in [Al/Mg] at [Fe/H]~$\approx -$1 could be achieved if LIMSs 
were efficient Al producers (cfr. Model~13 predictions for Na -- thick dotted 
line in Fig.~\ref{fig:naalmg}, panel a). Finally, it is seen that different 
models predict almost the same [Al/Mg] versus [Fe/H] behaviour in the solar 
neighbourhood (Fig.~\ref{fig:naalmg}, bottom panel), whereas the theoretical 
[Na/Mg] versus [Fe/H] relations may differ by as much as 0.5--1.0 dex or even 
more at the lowest metallicities (Fig.~\ref{fig:naalmg}, top panel). The best 
description of the observational [Na/Mg] versus [Fe/H] diagram would be 
obtained with a model assuming a set of yields for massive stars intermediate 
between the two sets used in Models~1 and 5, namely, \citet{ww95} case~B and 
\citet{k06} with $\varepsilon_{\mathrm{HN}}$~= 1, respectively (see 
Table~\ref{tab:nucp}).

\paragraph{\bf Silicon} 
Silicon is produced mostly during oxygen burning and its final abundance in the 
stellar ejecta is sensitive to a variety of factors regulating the evolution 
and explosion of the progenitor stars \citep{ww95}. Observationally, 
silicon abundances are affected by contamination from CH and H$\delta$ lines; 
departures from LTE do not seem to be a major concern, at least for Sun-like 
stars \citep{s08}.

At the lowest metallicities, where the signature of low-$Z$ SNeII is clearly 
seen, the [Si/Fe] ratio predicted by our GCE model may vary by up to 1~dex. 
Models~1 and 5, computed with \citeauthor{ww95}'s \citeyearpar{ww95} case~B and 
\citeauthor{k06}'s \citeyearpar{k06} $\varepsilon_{\mathrm{HN}}$~= 1 yields for 
massive stars, respectively, predict a mildly declining trend of [Si/Fe] versus 
[Fe/H] over the whole metallicity interval covered by the observations 
[Fig.~\ref{fig:alpha}, panel a, short-dashed (red) and lower solid (black) 
lines, respectively]. The theoretical trend is in good agreement with the 
available data, apart from the flattening traced by observations at supersolar 
metallicities \citep{b05}, which might require a revision of current stellar 
yields from high-metallicity massive stars and/or SNeIa. A very steep decline 
of the [Si/Fe] ratio with metallicity in the $-$4~$<$ [Fe/H]~$< -$1 metallicity 
range is obtained instead by assuming the yields of \citet{k06} with 
$\varepsilon_{\mathrm{HN}}$~= 0 [Model~4, Fig.~\ref{fig:alpha}, panel a, upper 
solid (blue) line], at odds with observations. This would indicate that in the 
very early halo phases most massive stars must explode as HNe rather than 
normal SNeII.

\paragraph{\bf Sulphur} 
Like silicon, sulphur is an intermediate-mass element whose production is 
directly attributable to oxygen burning, either at the center or in the 
convective shells of massive stars, or during explosive nucleosynthesis 
\citep[e.g.][]{ww95,lc03}. As a volatile element, it is hardly affected by 
depletion onto dust particles, which makes it extremely useful in deciphering 
the chemical enrichment of damped Lyman~$\alpha$ systems \citep[see 
e.g.][]{n04}. Hence, unraveling the history of S enrichment in the early Galaxy 
has become a fundamental step towards our understanding of the high-redshift 
universe. However, sulphur abundance determinations in Galactic stars are 
currently highly debated. \citet{t05} find non-LTE abundance corrections to be 
negligible for the $\lambda\lambda$~8693, 8694~\AA \ \ion{S}{I} lines of 
multiplet 6, while the $\lambda\lambda$~9212, 9228, 9237~\AA \ \ion{S}{I} lines 
of multiplet 1 suffer significant negative non-LTE corrections (up to 0.2--0.3 
dex). In the very metal-poor metallicity regime, a marked discordance is 
reported by \citet{t05} between the [S/Fe] values from the two abundance 
indicators: while the lines of multiplet 1 attain a nearly flat plateau (or 
even a slight downward bending), the lines of multiplet 6 show an 
ever-increasing trend with decreasing metallicity. A more recent analysis by 
\citet{n07}, however, based on high S/N ratio spectra of 40 metal-poor halo 
stars, using both weak ($\lambda$~8694.6) and strong ($\lambda\lambda$~9212.9, 
9237.5) \ion{S}{I} lines, and taking non-LTE corrections into account, finds a 
flat trend of [S/Fe] versus [Fe/H] and no stars with [S/Fe] in excess of 0.60 
dex.

In Fig.~\ref{fig:alpha}, panel b, the sulphur abundances measured by 
\citet[][crosses]{n07} are compared to the much more dispersed measurements by 
\citet[][triangles]{c05} from multiplets~1, 6 and 8. Disc data at higher 
metallicities are from \citet[][open squares]{r03}. Also shown in 
Fig.~\ref{fig:alpha}, panel b, are the predictions of our GCE models for the 
evolution of [S/Fe] versus [Fe/H] (solid and dashed lines). Models~1 and 5, 
adopting the yields of \citet{ww95} case~B and \citet{k06} with $\varepsilon$~= 
1 for massive stars, respectively, favour a nearly flat trend of [S/Fe] versus 
[Fe/H] in the metallicity regime $-$3~$<$ [Fe/H]~$< -$1. Model~4, instead, 
adopting the yields of \citet{k06} with $\varepsilon$~= 0 for massive stars, 
predicts a steeply decreasing trend of [S/Fe] versus [Fe/H] in the same 
metallicity range. It is worth noticing that the outcome from Model~4 is at 
variance with observations if the data by \citet{n07} are used for comparison, 
but becomes compatible with the observations if the data from other authors 
\citep[here][]{c05} are added. Clearly, more data on S abundances in metal-poor 
stars are needed to reduce the current uncertainties in GCE modelling of S 
evolution, and to rule out the set(s) of stellar yields which produce results 
at variance with the observations.

\paragraph{\bf Calcium} 
The production of calcium is directly attributable to incomplete silicon and 
oxygen burning \citep{ww95}. Departures from non-LTE in the line formation for 
both neutral and singly-ionized calcium have been recently investigated by 
\citet{mkp07}, for a wide range of metallicities of the stars. As for 
\ion{Ca}{I}, non-LTE abundance corrections depend on $T_\mathrm{eff}$, $\log 
g$, [Ca/H] and microturbulence value, and differ in value and sign for 
different \ion{Ca}{I} lines. For \ion{Ca}{II}, non-LTE leads to negative 
abundance corrections over the whole range of stellar parameters.

In Fig.~\ref{fig:alpha}, panel c, our model predictions on the behaviour of 
[Ca/Fe] as a function of [Fe/H] are compared to LTE abundances from 
\citet[][stars]{c04}, \citet[][pentagons]{l08}, \citet[][upside-down 
triangles]{g03}, \citet[][squares]{r03,rlap06} and \citet[][circles]{b05}. The 
uncertainties in the model predictions for Ca are of the same order of 
magnitude as those for Si and S. At very low metallicities, Models~1 and 5 
[short-dashed (red) line and lower solid (black) line, respectively] produce 
the best-fitting curves, whereas Model~4 [upper solid (blue) curve] 
overestimates the observed [Ca/Fe] ratios. None of the models is able to 
properly fit the thin-disc data.

\paragraph{\bf Titanium} 
The main isotope of titanium, $^{48}$Ti, is produced mainly during complete and 
incomplete silicon burning. Spectroscopists may rely on several lines for 
deriving Ti abundances in stars. To our knowledge there is only one published 
non-LTE study of Ti in cool dwarfs and giants by \citet{h97}; the data shown in 
Fig.~\ref{fig:alpha}, panel d, are all derived under the assumption of LTE.

As for K (see Sect.~\ref{sec:oddz}), also for Ti the theoretical predictions 
are offset with respect to the location of the data points in the [Ti/Fe] 
versus [Fe/H] diagram, though to a lesser extent (see Fig.~\ref{fig:alpha}, 
panel d). This clearly points to the need for a major revision of current SNII 
and HN yields, over the whole range of initial metallicities of the progenitor 
stars.

\subsection{Iron-group elements}
\label{sec:irong}

\paragraph{\bf Scandium} 
Scandium is produced in the innermost ejected layers of core-collapse SNe. It 
is made both as $^{45}$Sc during neon burning and as the radioactive progenitor 
$^{45}$Ti in explosive oxygen and silicon burning \citep{ww95}. It is now 
widely recognized that artificially induced explosions in the progenitor star 
models lead to incorrect nucleosynthesis predictions, a consideration which 
more generally applies to the composition of the whole innermost ejecta 
\citep[][and references therein]{fro06}. A quantity indispensable to correctly 
describe the nucleosynthesis in the innermost ejecta is the electron fraction 
in the layers undergoing explosive Si burning, $Y_e$. A value of $Y_e$ in 
excess of 0.5 is obtained with a consistent treatment of neutrino interactions. 
This brings about large amounts of Sc, Ti and Zn \citep[e.g.][]{fro06}, thus 
solving the problem of their underproduction of the values observed in 
low-metallicity stars. Alternatively, an enhanced Sc production is obtained 
with a low-density model \citep{un05}, where a weak jet expands the interior of 
the core-collapse SN progenitor before a strong jet develops, which explodes 
the star through a strong shock. Ti, Mg, Ca, Co and Zn production is enhanced 
in this model as well. The abundances of Sc, as well as Mn and Co, in the SN 
ejecta are then further enhanced by neutrino-nucleus interactions (the 
$\nu$-process) occurring in Si-burning regions \citep{yun08}. Unfortunately, 
these studies do not provide complete grids of stellar yields useful for GCE 
models.

\begin{figure*}
  \centering
  \includegraphics[width=\textwidth]{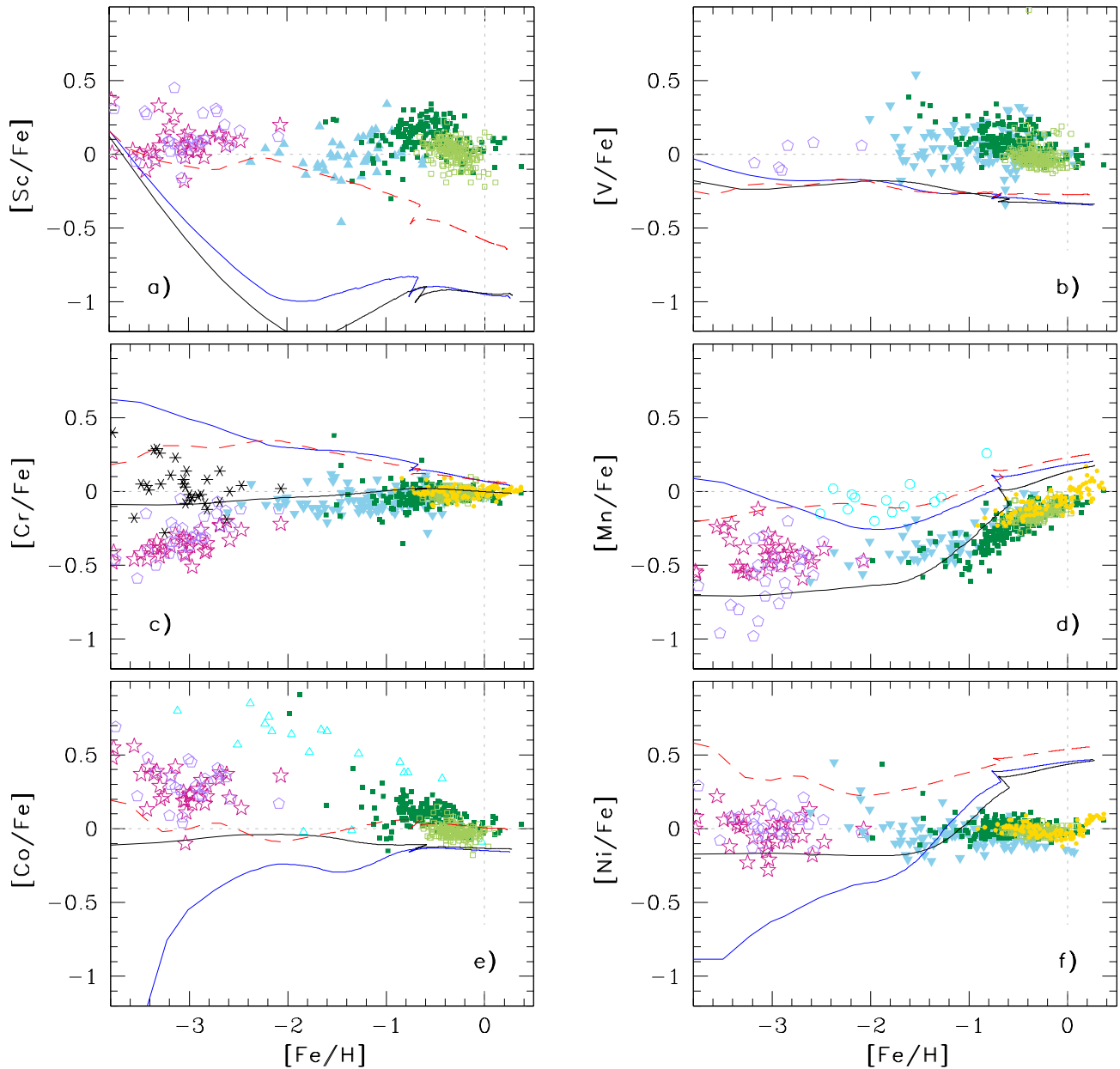}
  \caption{Predicted behaviour of the relative ratios of several Fe-peak 
    elements to iron (panel a: scandium, panel b: vanadium, panel c: chromium, 
    panel d: manganese, panel e: cobalt, panel f: nickel) with respect to the 
    relative iron abundance in the solar neighborhood, for Models~1 
    [short-dashed (red) lines], 4 [upper solid (blue) lines for Sc, V, Cr and 
    Mn; lower solid (blue) lines for Co and Ni] and 5 [lower solid (black) 
    lines for Sc, V, Cr and Mn; upper solid (black) lines for Co and Ni]. Also 
    shown are data from \citet[][stars]{c04}, \citet[][open pentagons]{l08}, 
    \citet[][asterisks, for Cr from \ion{Cr}{II} lines, see text]{b09}, 
    \citet[][open circles]{bg08}, \citet[][open triangles]{b10}, 
    \citet[][upside-down triangles]{g03}, \citet[][filled squares]{rlap06}, 
    \citet[][open squares]{r03}, \citet[][filled circles, only for Cr and 
    Ni]{b05} and \citet[][filled circles, only for Mn]{ffb07}. \citet{bg08} and 
    \citet{b10} performed non-LTE calculations; all the other abundance 
    studies, instead, rely on the assumption of LTE. See the electronic edition 
    of the journal for a colour version of this figure.}
  \label{fig:fepeak}
\end{figure*}

As a matter of fact, the predicted trend of [Sc/Fe] as a function of [Fe/H] 
severely disagrees with the [Sc/Fe] ratios measured in solar neighbourhood 
stars if the \citet{k06} yields for massive stars are used, independently of 
the assumed HN fraction (solid lines in Fig.~\ref{fig:fepeak}, panel~a). 
Marginal agreement between model predictions and observations is obtained by 
adopting the metallicity-dependent yields of \citet[][case B]{ww95} for massive 
stars, but only for [Fe/H]~$< -$1.5 [short-dashed (red) line in 
Fig.~\ref{fig:fepeak}, panel~a].

\paragraph{\bf Vanadium} 
Vanadium is synthesized mainly by incomplete explosive Si burning in massive 
stars \citep{lc03}.

The difficulty in measuring the V abundances in very metal-poor stars makes it 
troublesome to try and determine the trend of [V/Fe] versus [Fe/H] below 
[Fe/H]~$\approx -$1.5. It is clear though that current stellar yields 
underestimate the overall V abundance in the Galaxy, by at least 0.3--0.4~dex 
at disc metallicities (see Fig.~\ref{fig:fepeak}, panel~b). 

\paragraph{\bf Chromium} 
Chromium is made chiefly during incomplete explosive Si burning 
\citep{ww95,lc03}. Several authors \citep[e.g.][]{h04,l08,b09} have noticed 
that an offset exists between the abundances derived from \ion{Cr}{I} and 
\ion{Cr}{II} lines. Furthermore, an offset is found also between the abundances 
measured in giants and turnoff stars. The discrepancies may indicate non-LTE 
effects, unaccounted for in current analyses.

In Fig.~\ref{fig:fepeak}, panel c, we display the predicted behaviour of 
[Cr/Fe] versus [Fe/H] (curves), together with the data from different authors 
(symbols). Model~5 [lower solid (black) line], including HN nucleosynthesis, is 
the one in better agreement with the observations. It favours the abundances 
derived from \ion{Cr}{II} lines, rather than from \ion{Cr}{I}, as indicative of 
the Cr content of halo stars \citep[see also][]{k06}. Notice that \ion{Cr}{II} 
lines can only be observed in giants.

\paragraph{\bf Manganese} 
It is well established through theoretical modelling that manganese is mainly 
made in explosive silicon burning and nuclear statistical burning 
\citep[e.g.][]{a96}, but its yields from different types of SNe remain still 
uncertain. It is also debated whether or not the yields are metallicity 
dependent \citep{ffb07}. \citet{c08} have demonstrated that the evolution of Mn 
can be understood only if Mn yields from SNeIa are metallicity-dependent. This 
is a robust result, because it explains at the same time the evolution of Mn in 
the solar vicinity, the Galactic bulge and the Sagittarius dwarf spheroidal 
galaxy.

The Mn line at 602.1~nm is poorly reproduced by available hyper fine structure 
(HFS) linelist and should hence be discarded in careful abundance studies 
\citep{ffb07}. \citet{rlap06} include this line in their analysis, which may 
explain why they find similar trends of [Mn/Fe] versus [Fe/H] for their 
thick-disc and thin-disc samples, while \citet{ffb07} find different trends 
below [Fe/H]~= 0 (stars with kinematics typical of the thick disc show a steady 
increase of [Mn/Fe] with increasing [Fe/H], while stars with kinematics typical 
of the thin disc share almost the same [Mn/Fe] ratio). To our knowledge, there 
is only one work dealing with the important issue of non-LTE line formation for 
Mn \citep{bg08}. Though only 14 stars with [Fe/H] mostly from $-$1 to $-$2.5 
are analyzed, there is intriguing circumstantial evidence that non-LTE 
corrections might run up to 0.5--0.7~dex at low metallicity 
(Fig.~\ref{fig:fepeak}, panel d, open circles).

Let us now compare in Fig.~\ref{fig:fepeak}, panel d, our model predictions on 
the behaviour of [Mn/Fe] as a function of [Fe/H] (solid and dashed lines) to 
the data obtained under the assumption of LTE by \citet[][stars]{c04}, 
\citet[][open pentagons]{l08}, \citet[][upside-down triangles]{g03}, 
\citet[][squares]{r03,rlap06} and \citet[][filled circles]{ffb07}. At the low 
metallicities typical of halo stars, say [Fe/H]~$< -$1.5, the uncertainties in 
the GCE model predictions are very large, up to 0.8~dex. A model intermediate 
between Models~4 and 5 [upper (blue) and lower (black) solid lines, 
respectively], i.e. one allowing a significant fraction of high-mass stars to 
explode as HNe, can explain the observed [Mn/Fe] ratios very well. However, 
when considering the non-LTE corrections to the Mn abundances \citep[][open 
circles]{bg08}, the models without HNe (Models~1 and 4) better fit the data. At 
higher metallicities, and especially for [Fe/H]~$> -$0.8, all the models fail 
to reproduce the observations. This is due to the abrupt rise of the [Mn/Fe] 
ratio predicted owing to the late, overwhelming contribution to Mn synthesis 
from SNeIa. Compelling evidence for a reduction of the global Mn yield from 
SNeIa is discussed in \citet[][and references therein]{c08}. However, we feel 
that full non-LTE analysis of larger samples of stars is badly needed before we 
can draw any firm conclusion about the chemical evolution of Mn.

\paragraph{\bf Cobalt} 
Cobalt is synthesized by complete Si burning in the deepest stellar layers, 
whereas Cr and Mn form in the outer incomplete Si-burning regions. Up to a few 
years ago, it was common wisdom that the [Co/Fe] ratio as measured in halo 
stars steady increases with decreasing metallicity for $-$4.0~$<$ [Fe/H]~$< 
-$2.5, whereas the [Cr/Fe] and [Mn/Fe] ratios decrease with decreasing [Fe/H] 
over the same metallicity range in the same objects. The difference in the 
observed trends was first explained by the mass-dependent mass cut in SNeII by 
\citet{n99}. Then, \citet{un05} demonstrated that enhanced explosion energies 
may lead to higher Co and lower Cr and Mn production in low-metallicity, 
high-mass stars, in agreement with the observations. However, as mentioned in 
previous paragraphs, the usage of the \ion{Cr}{II} lines to estimate Cr 
abundances in giant halo stars and the correction of Mn abundances for non-LTE 
effects have led to a radically different picture, in which Cr and Mn stay 
almost constant rather than decline towards the lowest metallicity end. At the 
same time, investigations of the influence of non-LTE and HFS splitting on the 
formation of Co lines and abundances in cool stars have resulted in a trend of 
increasing [Co/Fe] with decreasing [Fe/H] much steeper than thought before 
\citep[][open triangles in Fig.~\ref{fig:fepeak}, panel~e]{b10}.

We have tried different prescriptions on the synthesis of Co in stars in the 
framework of our GCE model, but none of them has brought the model predictions 
into agreement with the relevant data (Fig.~\ref{fig:fepeak}, panel~e, solid 
and dashed lines). This is specially true if the non-LTE corrected values are 
used for comparison.

\paragraph{\bf Nickel} 
Nickel is mainly produced in the zones which undergo complete explosive Si 
burning in stars \citep[e.g.][]{lc03}. Measurements of Ni in solar 
neighbourhood stars indicate a flat trend of [Ni/Fe] versus [Fe/H] over the 
whole metallicity range (Fig.~\ref{fig:fepeak}, panel f, symbols).

None of our GCE models is able to explain the observed data 
(Fig.~\ref{fig:fepeak}, panel f, solid and dashed curves). The uncertainties in 
the GCE model predictions are particularly severe at the lowest metallicities, 
where the signature of SNII nucleosynthesis clearly shows up. Partial agreement 
between theoretical predictions and observations is found with Model~5, 
adopting \citet{k06} yields for massive stars with $\varepsilon$~= 1, but only 
in the low metallicity domain. The jump in [Ni/Fe] seen at [Fe/H]~$\sim -$1 is 
due to Ni production from SNeIa \citep[see also][]{k06}. The Ni yield from 
SNeIa is a strong function of $Y_e$ in the burning region, which, in turn, is 
determined by electron captures and, hence, is sensitive to uncertain 
quantities such as the propagation speed of the burning front and the central 
density of the white dwarf progenitor \citep{i99}. This leaves room for a 
downward revision of current Ni yields from SNeIa \citep{k06}.

\begin{figure}[b!]
  \centering
  \includegraphics[width=\columnwidth]{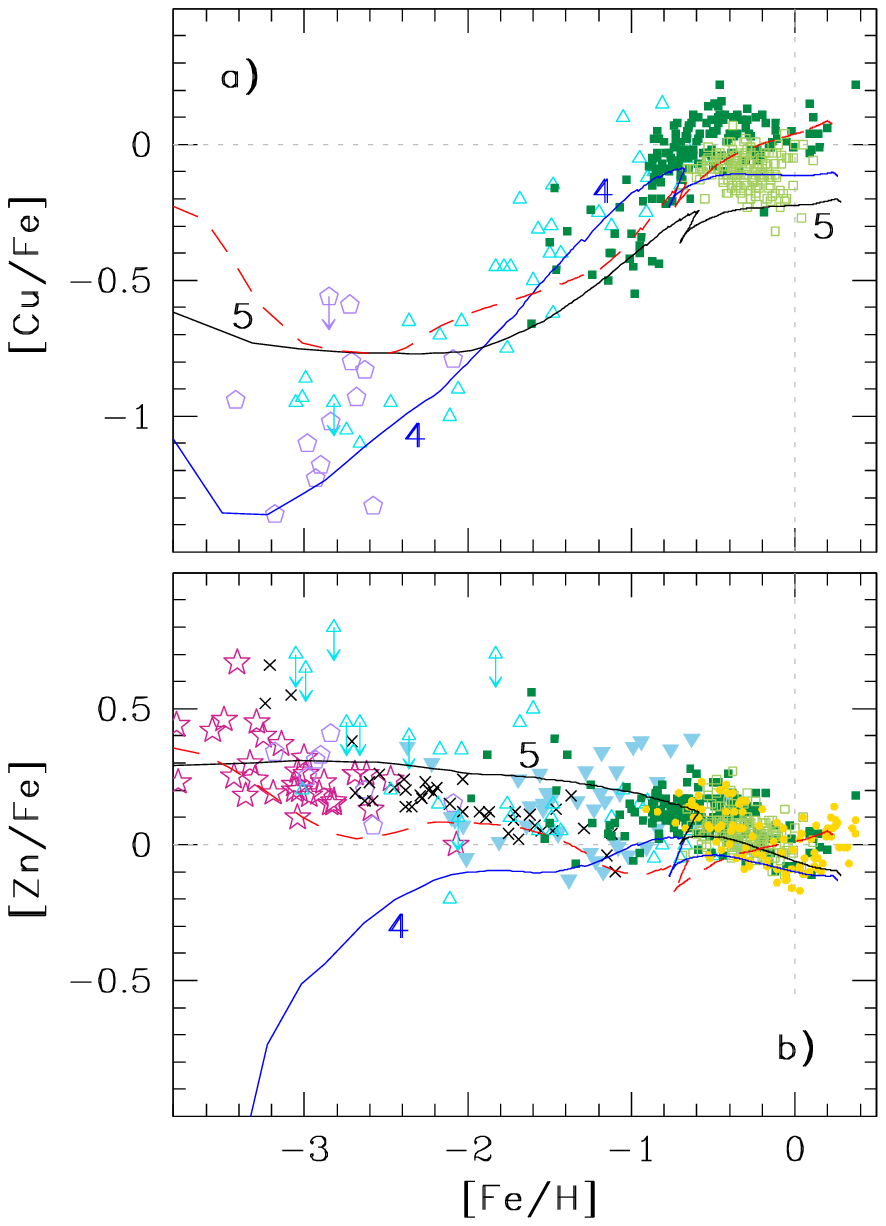}
  \caption{ [Cu/Fe] versus [Fe/H] (upper panel) and [Zn/Fe] versus [Fe/H] 
    (lower panel) relations as predicted by Models~1 [short-dashed (red) 
    lines], 4 and 5 (solid lines labelled 4 and 5, respectively) for the solar 
    neighbourhood. Also shown are data from \citet[][stars]{c04}, \citet[][open 
    pentagons]{l08}, \citet[][crosses]{n07}, \citet[][open triangles]{bih04}, 
    \citet[][upside-down triangles]{g03}, \citet[][filled squares]{rlap06}, 
    \citet[][open squares]{r03} and \citet[][filled circles]{b05}. See text for 
    discussion.}
  \label{fig:cuzn}
\end{figure}

\paragraph{\bf Copper} 
\citet{rm07} have recently presented a comprehensive study of copper evolution 
in different systems. They have demonstrated that, on a galactic scale, Cu is 
mainly produced by massive stars, rather than by SNeIa as previously thought 
\citep{m93,m02}. In massive stars, Cu is made during core-helium and 
carbon-shell hydrostatic burnings, as well as in explosive complete Ne burning 
\citep[e.g.][]{ww95,lc03}.

In Fig.~\ref{fig:cuzn}, panel a, we compare the predictions from Models~1 
[short-dashed (red) line], 4 [solid (blue) line labelled 4] and 5 [solid 
(black) line labelled 5] to Cu data from a number of studies spanning the whole 
metallicity range from [Fe/H]~$\approx -$4 to [Fe/H]~$\approx +$0.5. In 
low-metallicity stars, Cu abundances are derived under the assumption of LTE 
from the near-UV lines of \ion{Cu}{I} at 3247.53 and 3273.95~\AA, taking the 
effects of HFS and isotopic splitting into account \citep{bih04,l08}. At higher 
metallicities, Cu abundances \citep{r03,rlap06} are obtained from LTE analysis 
of optical \ion{Cu}{I} lines (5105.550, 5218.210 and 5220.090~\AA); thorough 
consideration of the HFS and isotopic splitting is required for the 5105~\AA \ 
line. As already pointed out by \citet{rm07}, the Cu content of extremely 
metal-poor stars is basically determined by explosive nucleosynthesis in 
massive stars, but as the metallicity grows, the contribution from the weak 
$s$-process operating in massive stars becomes increasingly important.

The uncertainties in modelling the Galactic evolution of Cu range from 0.9~dex 
at the lowest metallicities to 0.3~dex or even less above [Fe/H]~= $-$2.5. 
Apparently, Model~1, adopting \citet{ww95} case~B yields for normal SNeII, 
provides the best fit to the observed [Cu/Fe] versus [Fe/H] trend for 
[Fe/H]~$> -$2. At lower metallicities, a model intermediate between Models~4 
and 5, accounting for a fraction of HNe, reproduces better the data. However, 
the reader should be aware that Cu formation through the main $s$-process 
acting in low-mass AGB stars has not been included in the models, following the 
results of stellar nucleosynthesis studies that found it to be negligible 
\citep[see][and references therein]{rm07}. We feel that, before drawing any 
firm conclusion on which yield set would be better for Cu, it would be worth 
re-investigating the issue of Cu production in LIMSs by means of updated AGB 
models. Low-metallicity intermediate-mass AGB stars could in principle produce 
enough Cu to affect the chemical evolution of this element. For example, a 
5-M$_{\sun}$ AGB star with [Fe/H]~$\approx -$2.3 produces [Cu/Fe]~$\sim$ 
1.0~dex. Cu is produced by neutrons released during convective thermal pulses 
by the $^{22}$Ne($\alpha$, n)$^{25}$Mg reaction, similar to the weak 
$s$-process in massive stars. At present, yields of Cu from a full range of AGB 
masses and metallicities are not available to test this with a chemical 
evolution model.

\paragraph{\bf Zinc}
Similarly to sulphur, zinc is seen in damped Lyman $\alpha$ systems where, due 
to its refractory behaviour, constitutes a sensible nucleosynthetic tracer 
\citep[][and references therein]{n04}. Observations of stars in the solar 
vicinity show that [Zn/Fe] is close to zero for metallicities in the interval 
$-$2.0~$<$ [Fe/H]~$<$ +0.4 \citep[e.g.][]{b05,n07}, but rises steeply to 
[Zn/Fe]~$\approx$ +0.5 at the lowest metallicities \citep{c04}. Such a trend is 
reinforced when non-LTE corrections are applied \citep{t05}. An extreme 
alpha-rich freezeout with $Y_e >$ 0.5 is a practical mechanism to make zinc in 
high abundance from metal-poor SNeII \citep{fro06}. An alternative possibility 
for high Zn/Fe production are Pop\,III core-collapse very massive ($m \sim$ 
500--1000 M$_{\sun}$) stars, in which, at the presupernova stage, 
silicon-burning regions occupy a large fraction, more than 20 per cent of the 
total mass \citep{o06}.

In Fig.~\ref{fig:cuzn}, panel b, we compare the predictions from Models~1 
[short-dashed (red) line], 4 and 5 (solid lines labelled 4 and 5, respectively) 
to LTE Zn abundances from several studies (symbols; see figure caption). Large 
uncertainties are associated to the model predictions at metallicities typical 
of the halo phase. None of the models is able to reproduce the observations: 
Model~1 is in qualitative agreement with the observed trend of increasing 
[Zn/Fe] with decreasing [Fe/H] in the metallicity interval $-$4~$<$ [Fe/H]~$< 
-$1, but it underpredicts Zn. In contrast, the [Zn/Fe] versus [Fe/H] relation 
predicted by Model~5 for [Fe/H]~$< -$1.5 is flat. Finally, Model~4 produces by 
far less Zn than required to explain the observations of this element in solar 
neighbourhood stars.

\section{The Galactic gradient}
\label{sec:grad}

\begin{figure*}
  \centering
  \includegraphics[width=\textwidth]{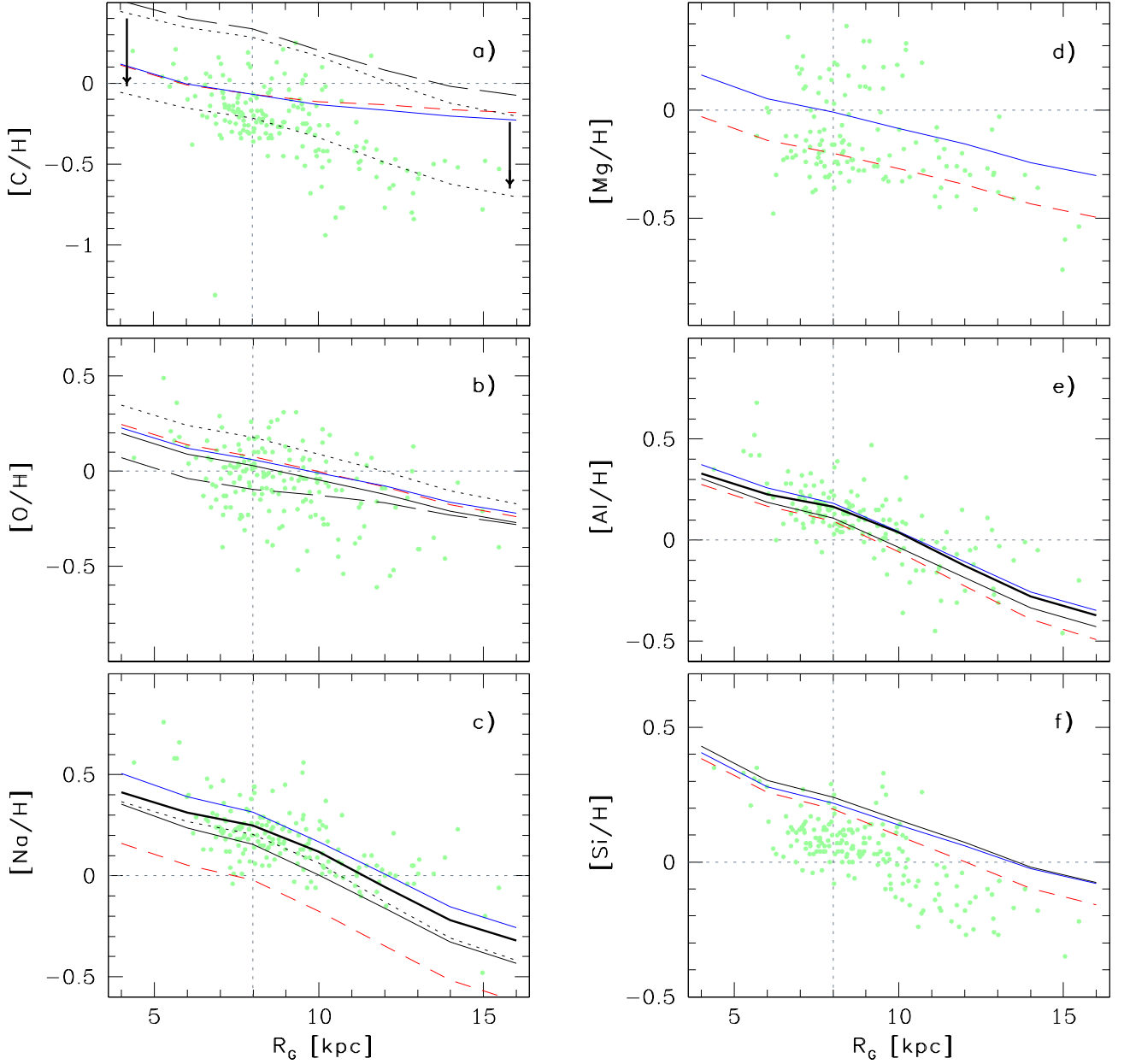}
  \caption{ Radial distribution of the abundances of carbon, oxygen, sodium, 
    magnesium, aluminium and silicon in the disc of the Milky Way at the 
    present time. The predictions from Models~1 [short-dashed (red) lines], 3 
    [long-dashed (black) lines, only for C and O], 4 [upper solid (blue) 
    curves], 5 [lower solid (black) curves], 9 [dotted (black) lines, only for 
    C, O and Na] and 15 (thick solid lines, only for Na and Al) are compared to 
    the homogeneous data for Cepheid variable stars by \citet{a02b,a02a,a02c}, 
    \citet{l03}, \citet{a04}, \citet{k05} and \citet{l06}. In panels~a and d 
    the predictions from Model~5 are not shown because they are pretty much the 
    same as Model~4. The intersection of the short-dashed vertical and 
    horizontal (grey) lines in each panel marks the position of the Sun. See 
    text for details. See the electronic edition of the journal for a colour 
    version of this figure.}
  \label{fig:grad}
\end{figure*}

The distribution with Galactocentric distance of the abundances predicted by a 
GCE model depends not only on the adopted yields, but even more on the 
assumptions on the galactic parameters, star formation and infall laws and IMF. 
In particular, the slope of the gradients is strictly related to the radial 
variation of the star formation/infall rates, i.e. of the enrichment/dilution 
rates \citep[e.g.][]{t88,mf89}. The yields, however, provide the absolute 
abundances and are therefore important ingredients for the gradients too.

The intriguing theme of the establishment and evolution of abundance gradients 
will be the subject of the next paper in this series. Here, for the sake of 
completeness we simply comment on the predicted present-day gradient for a few 
elements of interest. In Fig.~\ref{fig:grad}, we show the present-day gradients 
of carbon, oxygen, sodium, magnesium, aluminium and silicon predicted by 
Models~1, 3, 4, 5, 9 and 15. For the comparison with the observations, we use 
the homogeneous data for nearly 200 classical Cepheid variable stars presented 
in a series of papers by S.~Andrievsky and collaborators 
\citep{a02b,a02a,a02c,l03,a04,k05,l06}.

\begin{figure*}
  \centering
  \includegraphics[width=\textwidth]{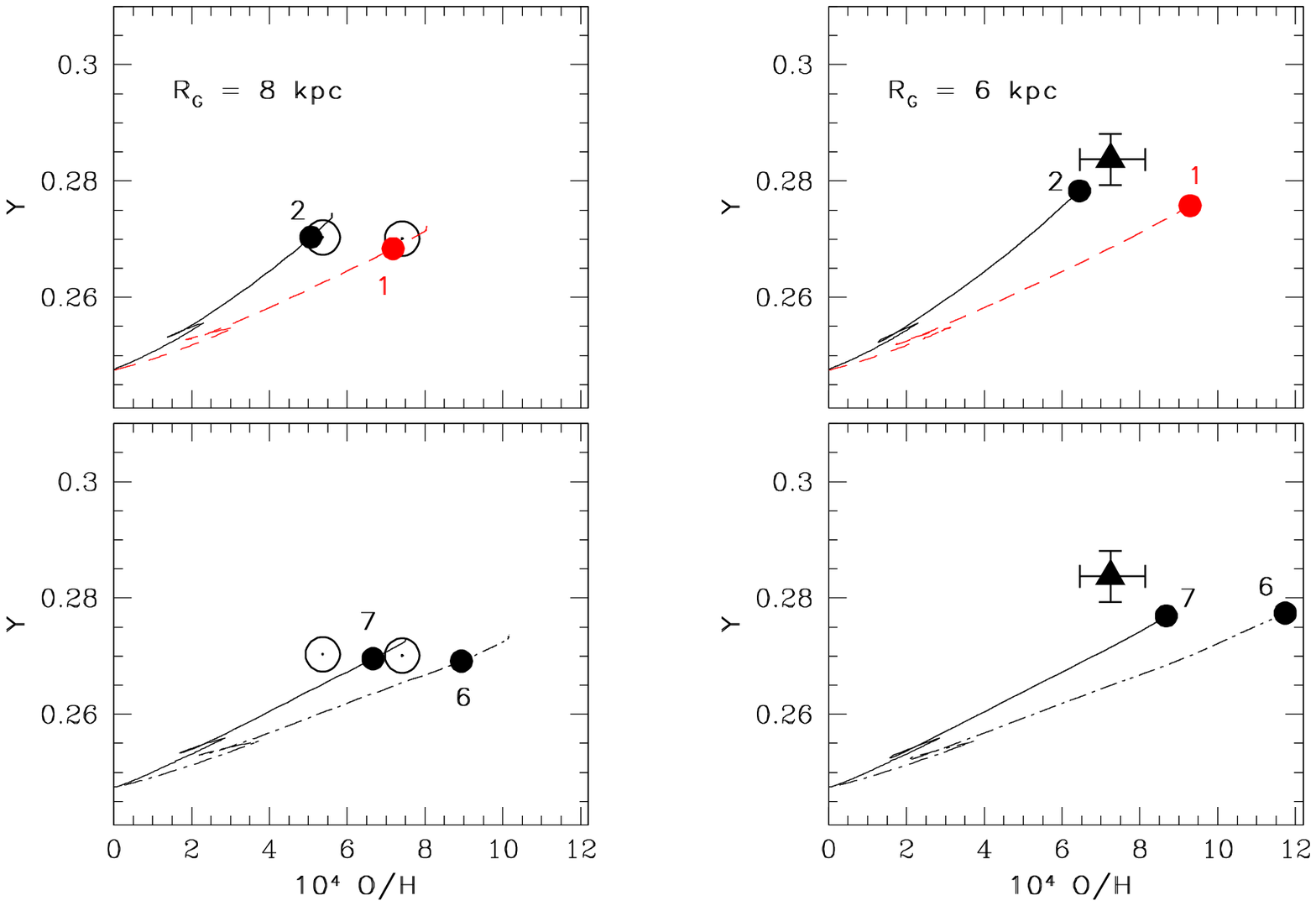}
  \caption{ $Y$ versus 10$^4$\,O/H in the solar vicinity (left panels) and in 
    the inner disc ($R_{\mathrm{G}}$~= 6~kpc; right panels), as predicted by 
    models with different nucleosynthesis prescriptions. Top panels: Model~1 
    [short-dashed (red) lines] versus Model~2 [solid (black) lines], showing 
    the effect of changing the mass cut in models of SNII explosions. Bottom 
    panels: Model~6 [dot-dashed (black) lines] versus Model~7 [solid (black) 
    lines], with and without rotation in high-mass stars, respectively. Left 
    panels: the filled circles on each theoretical track indicate the $Y$ and 
    10$^4$\,O/H values predicted at the time of the birth of the Sun, 4.5~Gyr 
    ago. These are compared with the bulk (initial) abundances of He and O in 
    the Sun (Sun symbols, from \citealt{gs98} -- higher O abundance -- and 
    \citealt{a09} -- lower O abundance). The effects of diffusion are taken 
    into account in defining the protosolar abundances displayed in the figure 
    \citep[see the discussion in][]{a09}. Right panels: the final $Y$ and 
    10$^4$\,O/H values attained by the models are highlighted (filled circles) 
    and compared to the abundances of He and O measured for M\,17, an 
    \ion{H}{II} region located at 2.1~kpc distance from the Sun (filled 
    triangles; see text for discussion).}
  \label{fig:he1}
\end{figure*}

It is clear that for some elements different choices of the yields do not 
provide only different \emph{absolute abundances,} but also different 
\emph{slopes} of the gradients. This is the case for carbon and oxygen, while a 
negligible effect is seen for the other elements, including elements heavier 
than Si, not shown in Fig.~\ref{fig:grad}. It is worth noticing, in particular, 
that the models including mass loss and rotation in high-mass stars are the 
ones that better reproduce the slope of the present-day C gradient. As an 
example, in Fig.~\ref{fig:grad}, panel a, we show the predictions of Model~9, 
using the Geneva group yields for rotating massive stars (upper dotted line). 
The theoretical curve is arbitrarily shifted by $-$0.5 dex (lower dotted line) 
to better compare the slope of the theoretical gradient with that of the 
gradient observed in Cepheids. Other models, using different nucleosynthesis 
prescriptions, predict a much flatter C gradient, at variance with 
observations. Model~3, using the older \citet{m92} yields for solar-metallicity 
massive stars computed with high mass loss but no rotation, is in agreement 
with the slope of the observed C gradient, but produces a too flat O gradient 
[Fig.~\ref{fig:grad}, panels~a and b, long-dashed (black) lines].

There are basically two conditions that the models must meet to be judged in 
agreement with the observations. First, they have to pass slightly above the 
intercept of the dashed lines indicating the position of the Sun in each 
diagram. Second, they have to reproduce the slope of the observed gradient. The 
intersection of the vertical and horizontal dashed lines marks the position of 
the Sun in each panel. If the composition of the Sun is representative of the 
typical chemical enrichment occurring in the Galactic disc at a distance of 
$R_{\mathrm{G}}$~= $R_{\mathrm{G}, \sun}$~= 8~kpc from the Galactic centre, one 
expects the abundances of Cepheids at comparable Galactocentric distances to 
scatter slightly above the solar value, under the assumption that the 
metallicity of the ISM has only slightly evolved from the time of the Sun 
formation up to now \citep{t88}. This is what is found for most elements, with 
the notable exceptions of carbon \citep[on average lower than solar in 
Cepheids, because of the occurrence of the first dredge-up;][]{k05}, sodium and 
aluminium \citep[on average enhanced with respect to solar, again reflecting 
evolutionary processes inside the stars themselves;][]{k05}. While this may 
complicate the comparison with the absolute abundances of C, Na and Al 
predicted by the models, our conclusions on the slopes of the gradients should 
remain basically unaffected.

\begin{figure*}
  \centering
  \includegraphics[width=\textwidth]{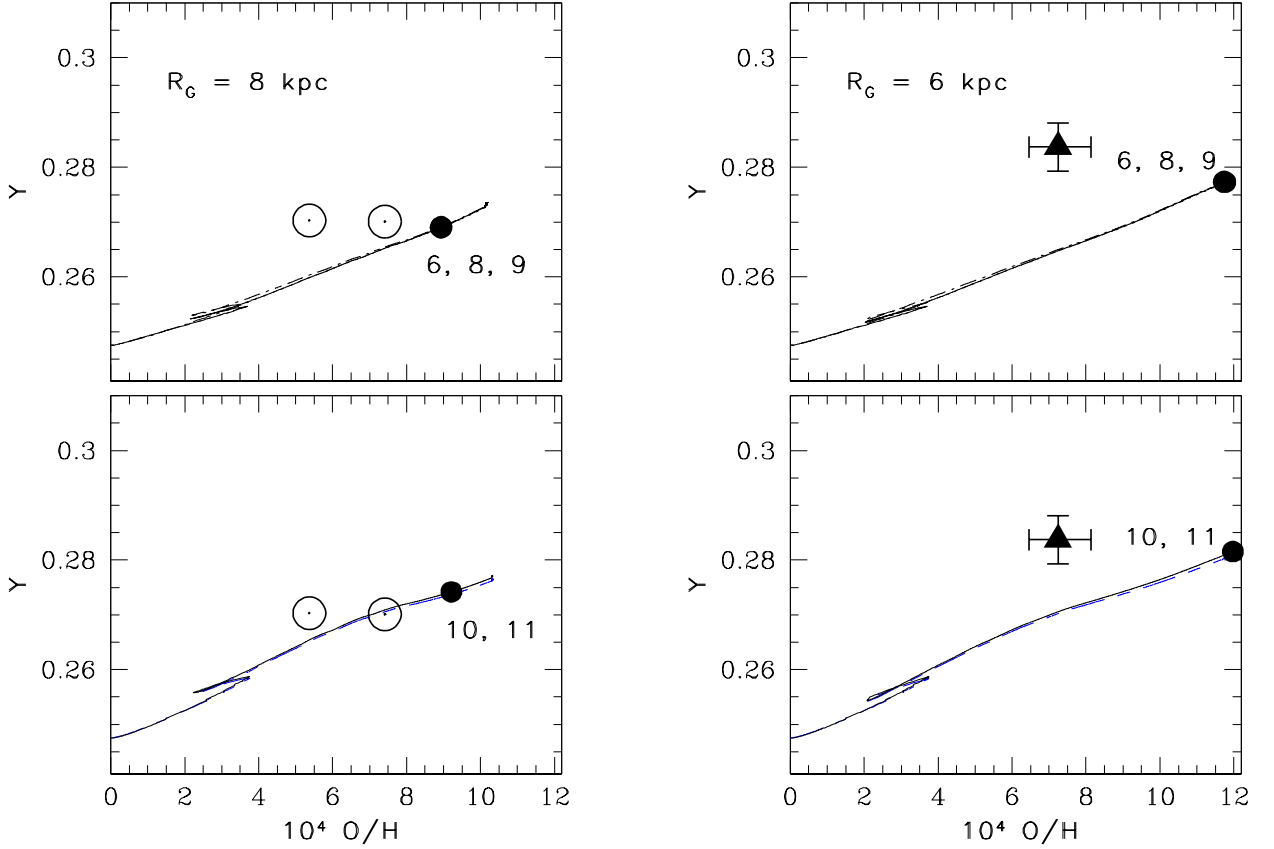}
  \caption{ Same as Fig.~\ref{fig:he1}, but confronting the predictions from 
    Models~6, 8 and 9 (upper panels; dot-dashed, solid and dotted lines, 
    respectively) and 10 and 11 (lower panels; short-dashed-long-dashed and 
    solid lines, respectively). The different model predictions overlap. This 
    suggests that the uncertainties associated with the treatment of mass loss 
    and HBB in AGB do not have an impact on the predicted $Y$ versus 
    10$^4$\,O/H behaviour.}
  \label{fig:he2}
\end{figure*}

In conclusion, we have seen that, at least for some elements, the comparison of 
the gradients predicted by using different grids of yields with a homogeneous 
data set can prove useful to discriminate among different nucleosynthesis 
prescriptions.

\section{The helium-to-heavy element enrichment ratio}
\label{sec:dydz}

\begin{figure*}
  \centering
  \includegraphics[width=\textwidth]{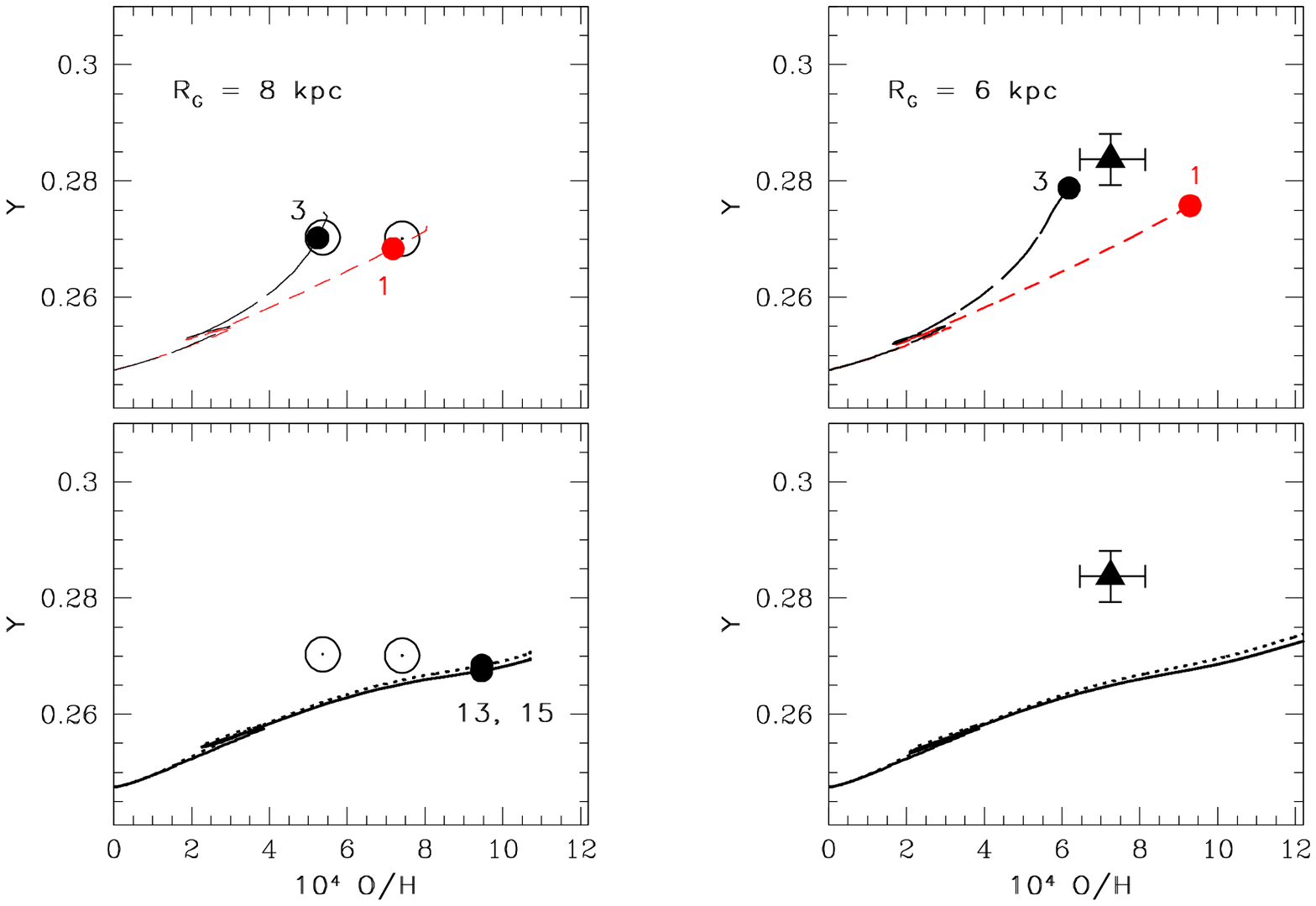}
  \caption{ Same as Fig.~\ref{fig:he1}, but confronting the predictions from 
    Models~1 and 3 [upper panels; short-dashed (red) and long-dashed (black) 
    lines, respectively] and 13 and 15 (lower panels; thick dotted and thick 
    solid lines, respectively).}
  \label{fig:he3}
\end{figure*}
%

\begin{figure}
  \centering
  \includegraphics[width=\columnwidth]{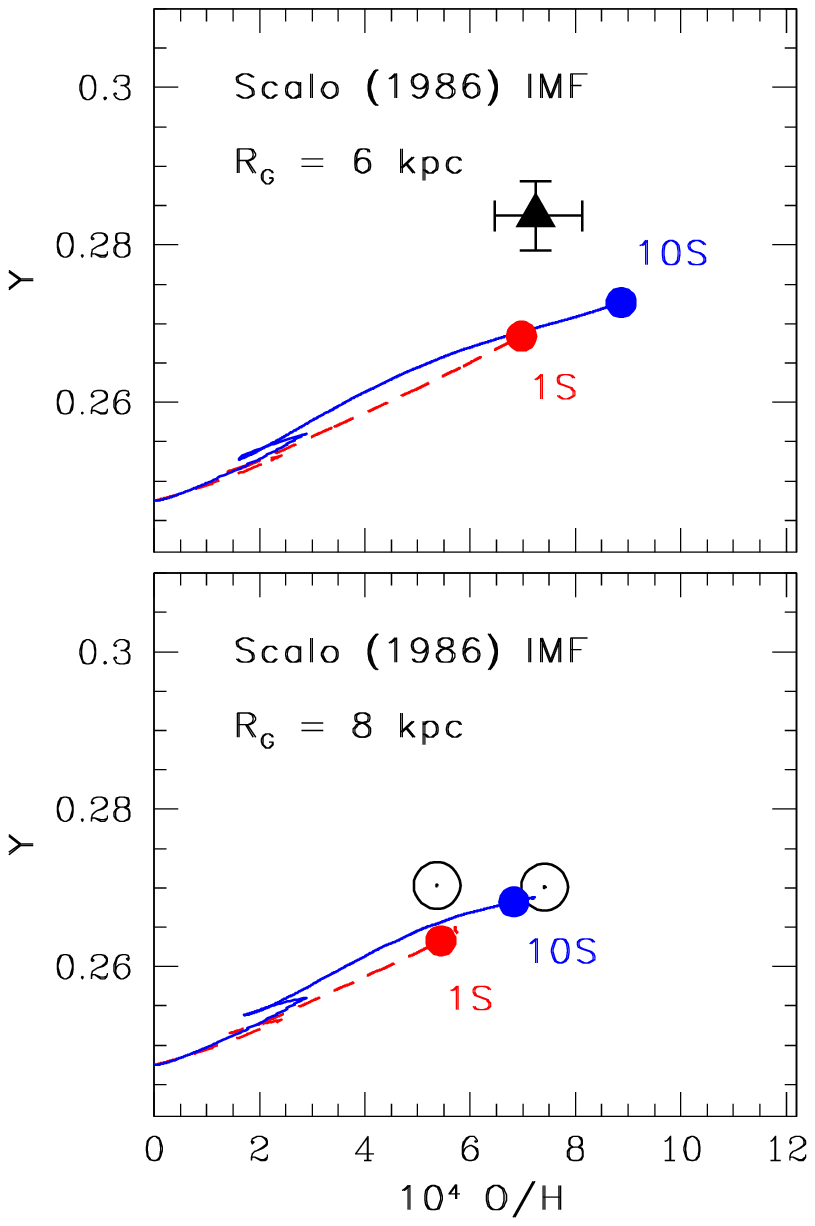}
  \caption{ $Y$ versus 10$^4$\,O/H predicted by models for the inner disc 
    ($R_{\mathrm{G}}$ = 6~kpc; top panel) and for the solar vicinity (bottom 
    panel) with different nucleosynthesis prescriptions [short-dashed (red) 
    lines: Model~1S; solid (blue) lines: Model~10S]. The models adopt a 
    \citet{s86} IMF rather than a \citet{k93} IMF. `Observational' values for 
    M\,17 and the Sun are shown as well (see text).}
  \label{fig:heScalo}
\end{figure}

In Figs.~\ref{fig:he1} to \ref{fig:he3} we show the predicted behaviour of the 
helium mass fraction, $Y$, versus 10$^4$\,O/H in the ISM for two Galactic 
radii, $R_{\mathrm G} = R_{\mathrm{G, \sun}}$~= 8~kpc (left panels) and 
$R_{\mathrm G}$~= 6~kpc (right panels). Shown are the predictions of models 
computed with different nucleosynthesis prescriptions (see Figs.~\ref{fig:he1} 
to \ref{fig:he3} captions and detailed description of the adopted 
nucleosynthesis in Table~\ref{tab:nucp}).

The predictions of GCE models on the composition of the ISM in the solar 
vicinity 4.5 Gyr ago are usually compared, for elements heavier than He, to the 
photospheric solar abundances. However, this is incorrect, since the 
photospheric abundances are no longer representative of the original 
composition of the Sun at birth, because of the effects of diffusion 
\citep[e.g.][]{bsb06}. The solar abundances of He and O displayed in 
Figs.~\ref{fig:he1} to \ref{fig:he3} for comparison with the theoretical model 
results (Sun symbols) have been corrected for the effects of diffusion 
\citep{a09}. We show both the values recommended by \citet{gs98}, 
$Y_{\sun,\, \mathrm{ini}}$~= 0.2701, 
log\,$\varepsilon$(O)$_{\sun,\, \mathrm{ini}}$~= 8.87 
[log\,$\varepsilon$(O)$_{\sun,\, \mathrm{phot}}$~= 8.83 $\pm$ 0.06], and those 
recommended by \citet{a09}, $Y_{\sun,\, \mathrm{ini}}$~= 0.2703, 
log\,$\varepsilon$(O)$_{\sun,\, \mathrm{ini}}$~= 8.73 
[log\,$\varepsilon$(O)$_{\sun,\, \mathrm{phot}}$~= 8.69 $\pm$ 0.05].

It can be immediately seen that the largest uncertainties in the model 
predictions are associated to the predicted oxygen abundances, while all the 
models predict virtually the same $Y$ value at $R_{\mathrm G}$~= 8~kpc at Sun's 
birth and at $R_{\mathrm G}$~= 6~kpc at the present time. Small differences are 
found in the model predictions when changing the prescriptions on He and O 
synthesis in LIMSs (Fig.~\ref{fig:he2} and lower panels of Fig.~\ref{fig:he3}). 
On the other hand, the differences become significant when changing the 
prescriptions on He and O synthesis in massive stars  (Fig.~\ref{fig:he1} and 
upper panels of Fig.~\ref{fig:he3}). In particular, the models adopting the 
yields by \citet{hmm05} for solar-metallicity rotating massive stars (Models~6, 
8, 9, 10, 11, 13 and 15) tend to overproduce oxygen in the Galaxy. In 
particular, the helium-to-metal enrichment ratio predicted by Model~13 around 
and above solar metallicity is $\Delta Y$/$\Delta Z \simeq$~1. This is barely 
consistent with the value of $\Delta Y$/$\Delta Z$~= 2.1 $\pm$ 0.9 suggested by 
\citet{casa07} for a large sample of nearby K dwarf stars. Models~3, 12 and 14 
(the latter not shown in the figures), adopting the older yields by \citet{m92} 
for solar-metallicity stars, instead, are consistent with 
$Y_{\sun,\, \mathrm{ini}}$~= 0.2703, 
log\,$\varepsilon$(O)$_{\sun,\, \mathrm{ini}}$~= 8.73, as suggested by 
\citet{a09}, and give $\Delta Y$/$\Delta Z$~= 1.3--1.5 for solar- and 
supersolar-metallicity stars, in good agreement with \citeauthor{casa07}'s 
\citeyearpar{casa07} findings.

The best way to discriminate between the \citet{hmm05} and \citet{m92} yields 
for solar-metallicity stars is to compute the chemical evolution at inner radii 
where, owing to the higher metallicities attained, the differences in the model 
predictions are expected to exacerbate. In the inner Galaxy, at a distance of 
2.1~$\pm$ 0.2~kpc from the Sun \citep{h08}, lies M\,17, the best Galactic 
\ion{H}{II} region to derive He and O abundances, owing to the smallest 
correction for the presence of neutral helium \citep[see][and references 
therein]{cp08}. In Fig.~\ref{fig:he3}, upper right panel, we compare the 
predictions from Models~1 and 3 for $R_{\mathrm G}$~= 6~kpc with the new 
determination of He and O abundances for M\,17 by \citet{cp08}. It is seen that 
the use of \citet{m92} yields for solar-metallicity massive stars improves the 
model predictions. Independent chemical evolution studies have already 
supported \citet{m92} yields for solar-metallicity massive stars 
\citep{gt95,car05,cp08,mc08}. On the other hand, it is worth noticing that also 
with the up-to-date yields computed by the Geneva group for \emph{non-rotating} 
massive stars, the fits to the $Y$, O/H values at $R_{\mathrm G}$~= 8~kpc at 
Sun's birth and at $R_{\mathrm G}$~= 6~kpc at the present time improve with 
respect to the case with rotation (see Fig.~\ref{fig:he1}, lower panels).

We deem it necessary to stress at this point that different assumptions about 
the model ingredients can change the conclusions. As an example, in 
Fig.~\ref{fig:heScalo} we show the results of Models~1S and 10S computed for 
$R_{\mathrm{G}}$~= 6~kpc (top panel) and $R_{\mathrm{G}}$~= 8~kpc (bottom panel), 
this time assuming a \citet{s86} IMF. Because of the highest mass fraction per 
stellar generation locked up in stars that do not contribute to the chemical 
enrichment and because of the lower mass fraction falling in the 8--100 
M$_{\sun}$ mass range according to this IMF, the models now produce less metals 
(and, hence, less oxygen) and less helium. Thus, while still consistent, within 
the uncertainties, with the initial He and O abundances in the Sun, they are by 
no means able to account for the abundances of He and O in M\,17. Adopting 
different nucleosynthesis prescriptions would only worsen the discrepancy 
between model predictions and observations.

Finally, it is worth noticing that no constraints can be set at present on the 
helium-to-metal enrichment ratio in the solar neighbourhood at ages older than 
that of the Sun \citep{casa07}. This means that no constraints can be set on 
the He stellar yields for $Z < Z_{\sun}$ from chemical evolution studies of the 
Galaxy.


\section{Summary and conclusions}
\label{sec:conc}

\begin{figure*}
  \centering
  \includegraphics[width=\textwidth]{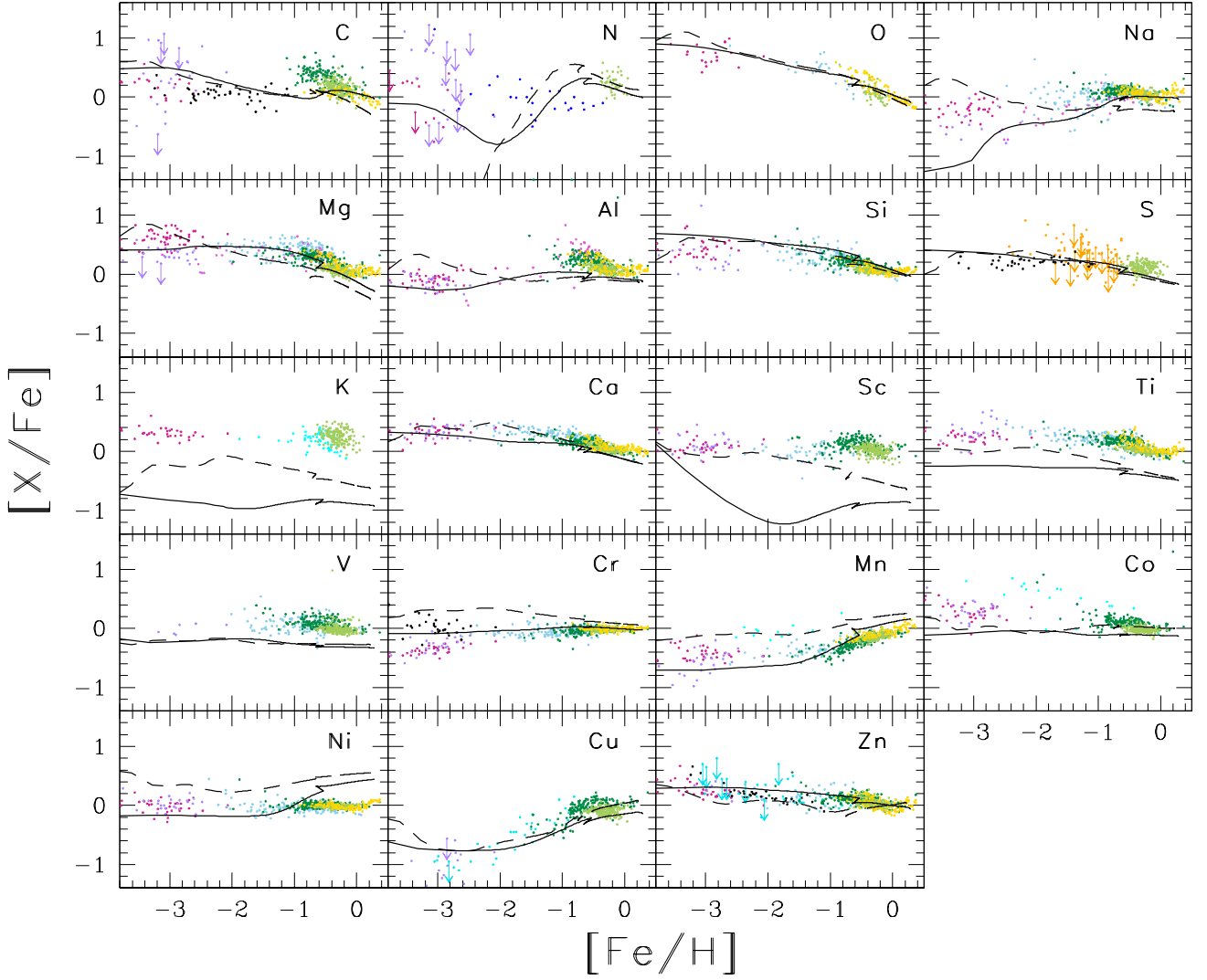}
  \caption{ [X/Fe] versus [Fe/H] relations for elements from C to Zn in the 
    solar neighbourhood. The predictions from Models~1 and 15 (dashed and solid 
    curves, respectively) are compared to data from several sources (see 
    captions to Figs.~\ref{fig:cn} to \ref{fig:cuzn} for references).}
  \label{fig:summ}
\end{figure*}
%

\begin{figure*}
  \centering
  \includegraphics[width=\textwidth]{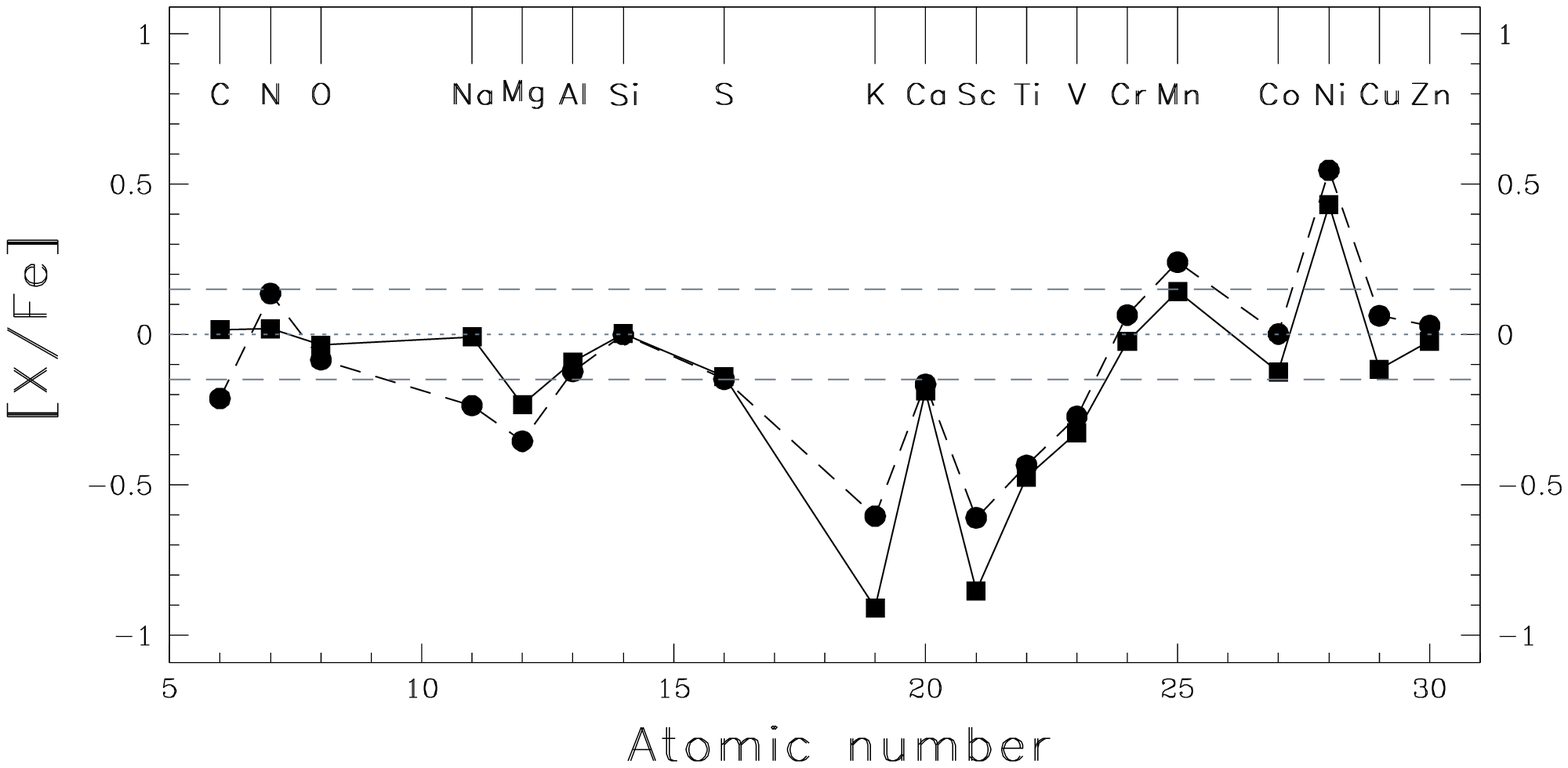}
  \caption{ [X/Fe] predicted by Model~1 (filled circles and dashed line) and 
    Model~15 (filled squares and solid line) at Sun's birth for elements from C 
    to Zn. The horizontal dotted line indicates the solar value of the ratios 
    that the models must reproduce. The horizontal dashed lines allow for a 
    $\pm$0.15 dex error in the model predictions.}
  \label{fig:solar}
\end{figure*}

In this paper, we have described the evolution of He and heavier elements from 
C to Zn in the Milky Way resulting from different sets of stellar yields and a 
well-tested chemical evolution model for the Galaxy.

In Sect.~\ref{sec:yields}, we have provided a detailed comparison among 
different yield sets suitable for use in chemical evolution models of the 
Galaxy. We stress that the absence of a complete and homogeneous grid of 
stellar yields, computed with the same input physics for many initial masses 
and chemical compositions of the stars -- say, from 0.8~M$_{\sun}$ to 
100~M$_{\sun}$ and from zero to super-solar metallicity -- is a major cause of 
uncertainty in the model predictions for elements such as He and N. In 
particular, detailed grids of yields for super-AGB stars are badly needed.

In Sect.~\ref{sec:model}, we introduced the adopted model for the chemical 
evolution of the Milky Way. This model well reproduces all the main observed 
features of the solar vicinity and of the whole Galactic disc. In particular, 
it is in agreement with the current star formation rate, gas and star radial 
density profiles and with the present-day Galactic SNII and SNIa rates and 
their ratio; it also reproduces the relative number of metal-poor-to-total 
stars, infall rate and degree of deuterium astration in the solar neighbourhood 
\citep[see][and \citetalias{r05}]{c01,r06}. All these quantities are reproduced 
\emph{independently of the choice of the stellar yields} and with a minimum 
number of free parameters, which justifies the use of the code to test possible 
stellar nucleosynthesis scenarios.

In Sects.~\ref{sec:abrat} to \ref{sec:dydz}, we described the model results, 
obtained by assuming different sets of stellar yields. In particular, in 
Sect.~\ref{sec:abrat} we dealt with the behaviour of several abundance ratios 
as functions of metallicity in the solar neighbourhood. The radial abundance 
profiles were discussed in Sect.~\ref{sec:grad}, whereas in 
Sect.~\ref{sec:dydz} we focused on the relative helium-to-metal enrichment in 
the Galaxy. A meaningful comparison between model predictions and observations 
relies on carefully selected, reliable data sets. Hence, in 
Sects.~\ref{sec:abrat} to \ref{sec:dydz}, we also tried to ascertain the origin 
of possible inconsistencies on different data sets. We find that, in many 
cases, non-LTE and/or 3D analyses of stellar spectra lead to a profound 
revision of the abundance trends inferred by means of 1D, plane-parallel model 
atmospheres under the LTE approximation. This deeply impacts on the 
interpretation of the model results.

Despite several considerable improvements in the field of stellar evolution and 
nucleosynthesis in recent years, no single combination of stellar yields is 
found which is able to reproduce at once all the available measurements of 
chemical abundances and abundance ratios in the Milky Way. In 
Figs.~\ref{fig:summ} and \ref{fig:solar}, we summarize the results of our `best 
yield choice', i.e. the combination of yields which maximizes the agreement 
with the largest number of observational constraints for the Milky Way. Our 
best model, Model~15 (solid lines in Figs.~\ref{fig:summ}--\ref{fig:solar}), 
uses the yields by \citet{k10} for LIMSs, the pre-supernova yields of He and 
CNO elements by the Geneva group for massive stars, and the results of 
explosive nucleosynthesis of \citet{k06} for SNeII below 20~M$_{\sun}$ and HNe 
above that limit. For sake of comparison, we also show the results of Model~1 
(dashed lines in Figs.~\ref{fig:summ}--\ref{fig:solar}), computed with the 
yields of \citet{vdhg} for LIMSs and \citet{ww95} for massive stars. This is 
the combination of stellar yields most often used in the literature.

Model~15 fits better the behaviour of [C/Fe], [N/Fe], [Na/Fe], [Mg/Fe], 
[Al/Fe], [Cr/Fe] and, perhaps, [Zn/Fe] as a function of [Fe/H] in the solar 
neighbourhood, although some caveats still remain:
\begin{enumerate} 
\item An exceptionally good fit between model predictions and observations of 
  carbon in the disc is obtained if the carbon abundances are those based on 
  the forbidden [\ion{C}{I}] line at 872.7~nm \citep[][golden points in 
  Fig.~\ref{fig:summ}]{bf06}; otherwise, the fit is rather poor.
\item The overall trend of [N/Fe] versus [Fe/H] can not be satisfactorily 
  explained by any model yet.
\item The behaviours of Na in halo stars and of Mg and Al at disc metallicities 
  are still far from being understood. Rotation might strongly impact on yields 
  of Na and Al from massive stars, but detailed computations suitable for use 
  in GCE models of the Galaxy are still missing in the literature.
\end{enumerate}
The evolution of oxygen in the solar neighbourhood and its solar abundance are 
very well reproduced by both models, thus indicating that its nucleosynthesis 
in stars is well understood by now. The same is true for Si, while S and Ca are 
slightly underestimated in the disc.

Potassium, scandium, titanium and vanadium are understimated by both models, 
with Model~15 performing worse for K, Sc and Ti. Ni is overestimated by 
Model~1, whereas Model~15 partially agrees with the data for halo stars.

Should future non-LTE analyses of larger samples of stars confirm the trends 
suggested by \citet{bg08} and \citet{b10} for Mn and Co, respectively, we would 
be led to conclude that: (i) none of the models is able to explain the Co data 
in the Galaxy; (ii) Model~1 reproduces the halo data for Mn better than 
Model~15; (iii) neither model can explain the behaviour of Mn in the Galactic 
disc, most likely because of an excessive synthesis of Mn from SNeIa at late 
times. Indeed, as demonstrated by \citet{c08}, who studied Mn evolution in the 
solar neighbourhood, the Galactic bulge and the Sagittarius dwarf spheroidal 
galaxy, considering metallicity-dependent yields of Mn from SNeIa could provide 
a solution to the problem of the overproduction of Mn at late times.

Finally, Model~1 provides a very good fit to copper abundances measured in 
solar neighbourhood stars, while Model~15 fails to reproduce Cu abundances in 
disc stars. However, the role of AGB stars as Cu factories has still to be 
assessed, which may lead to improvements in Model~15 predictions for Cu at disc 
metallicities.

While exploding core-collapse SNe through an artificially induced piston or 
thermal bomb is acceptable for the external layers, only a correct treatment of 
the physics of the explosion could provide correct yields for the species 
produced in the innermost ejected layers, such as the Fe-peak nuclei. Our 
results are in agreement with recent claims that a correct treatment of the 
explosion physics is still missing in SN models.

The inclusion of rotation in high-mass stellar models has allowed to sensibly 
improve the predictions of GCE models on C evolution in the early Galaxy. 
Contrary to common wisdom, high N/O ratios at the lowest metallicities, 
instead, can be obtained even if not allowing for low-metallicity massive stars 
to rotate fast.

Yields for LIMSs computed by taking rotation into account are provided in the 
literature, but stopping the computations at the beginning of the TP-AGB stage 
makes them useless in GCE models, since the important effects of the TDU and 
HBB on the final yields are not considered.

We sincerely hope that the results of this study can trigger a further effort 
to improve and expand the current existing grids of stellar yields.


\begin{acknowledgements}
      We thank the referee, Patrick Fran\c cois, for a thorough report which 
      helped us to improve the presentation of the paper. Thanks are due to 
      Monique Spite for providing the \ion{Cr}{II} data for halo giants of the 
      program ``First Stars''. DR and MT warmly thank Corinne Charbonnel for 
      the many enlightening conversations. DR's research at Bologna University 
      is supported by MIUR through grant PRIN~2007, prot.~no.~2007JJC53X\_001 
      (PI: F.~Matteucci). DR and MT acknowledge the hospitality of the 
      International Space Science Institute in Bern (CH) where some of the 
      described issues have been discussed during Team meetings.
\end{acknowledgements}

%
%

\bibliographystyle{aa}
\bibliography{RKTM10.bib}

\end{document}